\documentclass[aps,twocolumn,amsmath,amssymb,superscriptaddress,preprintnumbers]{revtex4-2}
\usepackage[pdftex]{graphicx}
\usepackage[outdir=./]{epstopdf}
\usepackage{enumerate,ulem,color}

\newcommand{\bk}{\boldsymbol{k}}
\newcommand{\bp}{\boldsymbol{p}}
\newcommand{\bq}{\boldsymbol{q}}
\newcommand{\br}{\boldsymbol{r}}
\newcommand{\bsig}{\boldsymbol{\sigma}}
\newcommand{\bv}{\boldsymbol{v}}
\newcommand{\Spin}{\mathcal{S}}
\newcommand{\bSpin}{\boldsymbol{\Spin}}
\newcommand{\bA}{\boldsymbol{A}}
\newcommand{\bB}{\boldsymbol{B}}
\newcommand{\bOm}{\boldsymbol{\Omega}}
\newcommand{\bmag}{\boldsymbol{m}}

\newcommand{\bnab}{\boldsymbol{\nabla}}

\newcommand{\FF}{\mathcal{F}}
\newcommand{\MM}{\mathcal{M}}
\newcommand{\Ham}{\mathcal{H}}

\newcommand{\ek}{\boldsymbol{e}_k}
\newcommand{\et}{\boldsymbol{e}_\theta}
\newcommand{\ep}{\boldsymbol{e}_\phi}

\newcommand{\vf}{v_\mathrm{F}}
\newcommand{\Tr}{\mathrm{Tr}}
\newcommand{\re}{\mathrm{Re}}

\newcommand{\modify}[1]{#1}

\begin{document}
\title{Spin-orbital magnetic response of relativistic fermions with band hybridization}

\author{Yasufumi~Araki}
% \email[]{\tt araki.yasufumi@jaea.go.jp}
\affiliation{Advanced Science Research Center, Japan Atomic Energy Agency (JAEA), Tokai 319-1195, Japan}

\author{Daiki~Suenaga}
% \email[]{\tt suenaga@mail.ccnu.edu.cn}
\affiliation{Research Center for Nuclear Physics (RCNP), Osaka University, Osaka 567-0047, Japan}

\author{Kei~Suzuki}
% \email[]{{\tt k.suzuki.2010@th.phys.titech.ac.jp}}
\affiliation{Advanced Science Research Center, Japan Atomic Energy Agency (JAEA), Tokai 319-1195, Japan}

\author{Shigehiro~Yasui}
% \email[]{\tt yasuis@keio.jp}
\affiliation{Research and Education Center for Natural Sciences, Keio University, Hiyoshi 4-1-1, Yokohama, Kanagawa 223-8521, Japan}
\affiliation{RIKEN iTHEMS, RIKEN, Wako 351-0198, Japan}

\begin{abstract}
    % !! description about orbital magnetic field?
    Spins of relativistic fermions are related to their orbital degrees of freedom.
    In order to quantify \modify{the effect of hybridization between relativistic and nonrelativistic degrees of freedom on spin-orbit coupling},
    we focus on the spin-orbital (SO) crossed susceptibility
    \modify{arising from spin-orbit coupling}.
    {The SO crossed susceptibility}
    is defined as the response function of their spin polarization
    {to the ``orbital'' magnetic field,
    namely {the effect of} magnetic field {on} the orbital motion of particles
    as the vector potential.}
    Once relativistic and nonrelativistic fermions are hybridized,
    their SO crossed susceptibility gets modified
    at the Fermi energy around the {band hybridization point},
    leading to spin polarization of nonrelativistic fermions as well.
    These effects are enhanced under a dynamical magnetic field
    that violates thermal equilibrium,
    arising from the interband process permitted by the band hybridization.
    Its experimental realization is discussed
    for Dirac electrons in solids with slight breaking of crystalline symmetry or doping,
    and {also for} quark matter {including} dilute heavy quarks strongly {hybridized} with light quarks,
    {arising in a relativistic heavy-ion collision process}.
\end{abstract}

\preprint{RIKEN-iTHEMS-Report-20}

\maketitle

\section{Introduction}
% Relativistic fermions arise at various energy scales.
% - elementary particles: quakrs at high energy.
% - low-energy excitations in Dirac semimetals.
% - (Weyl semimetals)
% - Relativistic effect: spin-momentum locking = spin-orbit coupling = helicity.
% Relativistic fermions sometimes coexist with nonrelativistic particles.
% - common in semimetals. (Co3Sn2Se2)
% - QCD Kondo effect.
% How to characterize the modulation of relativistic effect?
% Spin-orbital crossed susceptibility.
% - Definition of crossed susceptibility.
% - Relation to relativistic effect?
% Findings in this work.
% - Dynamical susceptibility.
% Arrangement.

Relativistic fermions arise at various energy scales.
While relativistic dynamics of fermions {is} generally described by the Dirac equation,
with four-component spinor field \cite{Dirac},
massless relativistic fermions can be described by the Weyl equation,
with two-component spinor field \cite{Weyl}.
Originally, those equations were invented
to describe elementary particles {obeying Lorentz symmetry} at high energy.
{Recently, they are also applied} to electrons in some crystalline materials,
classified as Dirac and Weyl semimetals, {which are} intensely studied over the past decade \cite{Murakami_2007,Wan_2011,Burkov_2011,Armitage_2018,Burkov_2018,Manna_2018,Araki_2019}.
In those semimetals, energy bands of electrons exhibit pointlike crossing structures in momentum space,
called Dirac or Weyl points,
around which the low-energy excitations of electrons can be effectively described as massless {Dirac or Weyl} fermions.
% Dirac and Weyl semimetals are distinguished by the degeneracies of the crossing points,
% corresponding to the number of components in Dirac/Weyl spinor,
% which is determined by the presence or absence of time-reversal and spatial inversion symmetries.
% For instance, Weyl points with broken time-reversal symmetry have been investigated in several magnetic compounds \cite{}.

% A Dirac point has fourfold degeneracy corresponding to the four-component structure of Dirac spinor,
% which usually comes from time-reversal and spatial inversion symmetries.
% A Weyl point, on the ohter hand, has twofold degeneracy due to breaking of either of these two symmetries,
% which arises from magnetism or the asymmetric crystalline structure.

While the characteristics of relativistic fermions themselves have been broadly studied,
we note here that relativistic fermions coexist with nonrelativistic fermions {in some cases}.
{It is found in some crystalline materials that}
Dirac or Weyl points coexist with other bands irrelevant to those point-node structures at the same energy \cite{Young_2012,Steinberg_2014,Chen_2014,Nakatsuji_2015,Kuroda_2017,Chang_2016,Ma_2019}.
Slight breaking of crystalline symmetries by lattice deformations or disorders
may lead to hybridization between the Dirac or Weyl bands and the irrelevant bands \cite{Yang_2014}.
For example, the magnetic alloy $\mathrm{Co_3 Sn_2 Se_2}$,
which is a sibling of the magnetic Weyl semimetal $\mathrm{Co_3 Sn_2 S_2}$,
is found to exhibit an anticrossing structure between the Weyl cones and the band irrelevant to them,
due to the strong band inversion by spin-orbit coupling (SOC) on Se \cite{Liu_2018,Xu_2018,Wang_2018}.
{Interband hybridization can occur in quark matter as well:}
light quarks, with flavors $u,d,$ or $s$, are usually treated as massless Dirac fermions,
in comparison with heavy quarks ($c$ and sometimes $b$).
If the heavy quarks are dilute enough,
they {form bound states with the light quarks at low momentum due to color exchange,}
proposed as the QCD Kondo effect \cite{Yasui_2013,Yasui_2016,Yasui_2017c,Yasui_2019c,Hattori_2015,Ozaki_2016,Yasui_2019,Yasui_2017,Kanazawa_2016,Kimura_2017,Yasui_2017a,Suzuki_2017,Yasui_2017b,Kimura_2019,Fariello_2019,Hattori_2019,Suenaga_2020,Suenaga_2020a,Kanazawa_2020,Araki_2021}.
Such a situation is proposed to occur at a short timescale after a relativistic heavy-ion collision process.

% !! distinguish spin and orbital effects.
Once nonrelativistic fermions are mixed and hybridized with relativistic fermions,
the relativistic effect, including SOC, may get modified.
In order to quantify the effect of the hybridization between relativistic and nonrelativistic degrees of freedom on SOC,
% Under the coexistence and hybridization of relativistic and nonrelativistic fermions as listed above,
% one needs a physical quantity that characterizes the relativistic effect present in the system.
we focus on the spin-orbital (SO) crossed susceptibility,
which constitutes a part of the magnetic susceptibility \cite{Tserkovnyak_2015,Koshino_2016,Nakai_2016,Suzuura_2016,Ando_2017,Ominato_2019,Aftab_2020,Ozaki_2021}.
The SO crossed susceptibility,
or the SO susceptibility in short,
is defined as the response function of \textit{spin} polarization (spin magnetization)
to the \textit{orbital} magnetic field,
namely, the effect of magnetic field {on} the orbital motion of particles via the vector potential.
% The idea of the SO susceptibility applies
% not only to electrons in solid states but also to particles at high energy.
The SO susceptibility arises from SOC,
namely, the correlation of spin and orbital degrees of freedom,
which is the major consequence of the relativistic effect \cite{Landau-Lifshitz}.
\modify{
    In connection with measurable transport properties,
    the SO susceptibility is related to the spin Hall conductivity \cite{Yang_2005,Murakami_2006},
    which is one of the typical transport properties arising from SOC \cite{Dyakonov_1971,Murakami_2003,Sinova_2015}.
}

In particular,
\modify{
    the characteristics of the SO susceptibility for relativistic (Dirac and Weyl) fermions
    have been intensely studied over the past few years,
    mainly in connection with topological insulators and semimetals \cite{Tserkovnyak_2015,Koshino_2016,Nakai_2016,Ominato_2019,Aftab_2020,Ozaki_2021}.
}
\modify{
    Since the spin of a massless relativistic fermion is locked to its momentum \cite{Schwartz},
    known as spin-momentum locking,
    it is proposed that the SO susceptibility of massless relativistic fermions shows a universal behavior,
    depending linearly on the Fermi energy (chemical potential) \cite{Koshino_2016,Nomura_2015}.
}
\modify{
    However, for multiband systems including nonrelativistic dispersions,
    general idea on the SO susceptibility has not been well established,
    despite its rising importance in both solid states and quark matter.
    Such a lack of general idea on the SO susceptibility is in a clear contrast to
    the situations in the spin-spin and orbital-orbital susceptibilities,
    which were generally formulated and studied for various kinds of materials
    from the mid-20th century \cite{Ashcroft,VanVleck,Blount_1962,Wannier_1964,Mishra_1969,Fukuyama_1970,Fukuyama_1971,Koshino_2007,Gomez-Santos_2011,Raoux_2015,Gao_2015}.
}

Based on the above background,
here we study the effect of band hybridization on the SO crossed susceptibility.
First, we derive a formula for the SO susceptibility applicable to general multiband systems.
Based on the obtained formula,
we evaluate the effect of hybridization.
\modify{
    In order to evaluate the difference in the SO susceptibility related to the presence or absence of the hybridization,
    we use a simple model composed of massless Dirac fermions obeying spin-momentum locking
    and nonrelativistic fermions free from SOC.
}
We find that,
if the magnetic field is suddenly switched on and violates the thermal equilibrium of the fermions,
which we call the dynamical \modify{process},
the susceptibility gets strongly reduced
at the Fermi level in the vicinity of the band hybridization point.
Owing to the band hybridization,
the nonrelativistic fermions \modify{acquire} spin polarization as well,
which also becomes significant \modify{in the dynamical process}.
% These modifications
% to the SO crossed responses
% arise from the interband {matrix elements emerging in the magnetic field perturbation},
% which is permitted {once the interband hybridization is included}.
% It 
\modify{
    We give a qualitative understanding of these modifications of the susceptibilities
    using the perturbation theory with a simple quantum mechanics,
    which we show in a similar manner with the well-established Van Vleck paramagnetism,
    namely the interband modification of spin-spin and orbital-orbital susceptibilities assisted by SOC \cite{VanVleck}.
}
% The modification in the spin polarization {found here}
% may be experimentally captured both in solid states and quark matter,
% which shall also be discussed in this article.

This article is organized as follows.
In Sec.~\ref{sec:formula},
we derive a general formula for the SO crossed susceptibility
by the linear response theory {using the Matsubara formalism}.
In Sec.~\ref{sec:weyl},
we apply the obtained formula to Weyl fermions as a test case,
to see the consistency with the previous literatures \cite{Koshino_2016,Nomura_2015}.
% On the basis of spin-momentum locking,
% we also discuss the relation between the SO susceptibility and the chiral magnetic effet (CME),
% which is another well-known characteristics of relativistic fermions
% in response to the orbital magnetic field \cite{Vilenkin_1980,Nielsen_1983,Kharzeev_2006,Kharzeev_2007,Kharzeev_2008,Fukushima_2008}.
{Section \ref{sec:model} is the main {study} in this article, where}
we introduce a minimal hybrid model with Dirac and nonrelativistic degrees of freedom,
and evaluate the SO crossed susceptibility using the obtained formula.
We discuss how the spin polarization in each sector,
namely Dirac or nonrelativistic,
gets modified by the hybridization.
In Sec.~\ref{sec:experiments},
we give some discussion on possible experimental methods to capture the obtained behavior of the susceptibilities,
both in solids and quark matter.
Finally, we summarize our analysis in Sec.~\ref{sec:conclusion}.
\modify{
    Detailed definitions of the susceptibilities and calculation processes are shown in the appendixes.
}
Throughout this article,
we take the natural unit with $\hbar =1$,
and the speed of light $c$ and the charge of particle $-e(<0)$ are left as constants.

\section{General analysis} \label{sec:formula}
In this section, we derive a general formula for the SO crossed susceptibility
% for general momentum-space Hamiltonian
by the linear response theory.
We start with the definition of the SO crossed susceptibility,
and evaluate it perturbatively by using the Matsubara Green's functions.
{After rearranging the obtained terms with the momentum-space (Bloch) eigenstates,}
{the SO susceptibility is expressed in terms of }
the geometric quantities related to the band eigenstates,
namely, the Berry connection, the Berry curvature, and the orbital magnetic moment.
% {as they couple to inhomogeneity in momentum space, which arise from the orbital magnetic field. }

\subsection{Linear response theory}
The SO crossed susceptibility is defined as the response function of \textit{spin} magnetization $\boldsymbol{M}^{\mathrm{s}}$
{to} the \textit{orbital} magnetic field $\boldsymbol{B}^{\mathrm{o}}$ \cite{Koshino_2016}.
\modify{
    We here give a brief discussion how the spin and orbital degrees of freedom are distinguished,
    by considering both electrons in solid states and elementary particles in high-energy physics,
    and show the definition of the SO crossed susceptibility.
    For detailed discussion about magnetic susceptibility among the spin and orbital degrees of freedom,
    see Appendix \ref{sec:orbital-spin}.
}

The spin magnetization is composed of the spin polarization of fermions,
\begin{align}
    \boldsymbol{M}^{\mathrm{s}} = -\gamma \langle \boldsymbol{S} \rangle, \label{eq:spin-magnetization}
\end{align}
where the coeffcient $\gamma = g \mu_B$ is the gyromagnetic ratio,
with $g$ the $g$-factor for the fermions and $\mu_B$ the Bohr magneton,
and $\boldsymbol{S}$ is the spin operator of the fermions.
{(Note that the spin magnetic moment of a negative-charge particle is antiparallel to the spin polarization.)}
The orbital magnetic field $\boldsymbol{B}^{\mathrm{o}}$ {is defined with} the U(1) vector potential $\bA$ satisfying $\boldsymbol{B}^{\mathrm{o}} = \bnab \times \bA$,
which couples to the particles in terms of the covariant derivative in continuum \cite{Weyl},
\begin{align}
    \bnab \mapsto \bnab - ie\bA,
\end{align}
and the Peierls phase on lattice \cite{Peierls_1933},
\begin{align}
    t_{ij} \mapsto t_{ij} \exp \left[-ie \int_{\br_i}^{\br_j} d\br \cdot \bA \right],
\end{align}
for the hopping amplitude between lattice sites $\br_i$ and $\br_j$.

{
    We should be careful about the role of $\boldsymbol{B}^{\mathrm{o}}$.
    In the context of relativistic quantum electrodynamics (QED),
    where the charged particles with Lorentz symmetry are coupled to the electromagnetic fields,
    the effect of magnetic field is fully described by the vector potential,
    namely, $\boldsymbol{B}^{\mathrm{o}}$ in our definition.
    $\boldsymbol{B}^{\mathrm{o}}$ couples to both the spin and orbital degrees of freedom in this framework.
    On the other hand, in the low-energy effective theory for nonrelativistic fermions,
    which is derived from the low-momentum expansion of massive Dirac fermions,
    $\boldsymbol{B}^{\mathrm{o}}$ does not fully describe the effect of the magnetic field.
    The Zeeman coupling, namely the direct coupling between the magnetic field and the spin angular momentum,
    is given separately from $\boldsymbol{B}^{\mathrm{o}}$ coupled as the vector potential.
    This framework applies to electrons in solid states,
    including those with Dirac or Weyl dispersion at low energy in their momentum-space band structures.
    In this framework,
    the effect of magnetic field on the magnetization via the Zeeman coupling,
    which is rather straightforward and has been well studied in the context of magnetism,
    is excluded from our analysis in response to $\boldsymbol{B}^{\mathrm{o}}$.
}
% Once we expand this theory for Dirac fermions with mass $m_D$ at low momentum,
% it gives the low-energy effective theory for nonrelativistic fermions,
% with the energy $p^2/2m_D$.
% The effect of magnetic field {splits into} the orbital effect as the vector potential,
% and the spin effect as the Zeeman coupling,
% namely the direct coupling between the magnetic field and the spin angular momentum.
% Therefore, in order to treat the effect of magnetic field on the fermions
% described by the nonrelativistic theory,
% we need to take into account the orbital effect and the Zeeman effect separately.
% In order to distinguish the Zeeman effect from the orbital effect in theoretical treatments,
% the magnetic field for the Zeeman effect is sometimes labeled as the spin magnetic field $\bB^{\mathrm{s}}$.
% Theoretically, the Zeeman effect of magnetic field is distinguished from the orbital magnetic field in the nonrelativistic regime,
% and sometimes labeled as the spin magnetic field $\bB^{\mathrm{s}}$.

With the above notations,
the SO crossed susceptibility is defined as a tensor $\chi^{\mathrm{so}}_{ij} \ (i,j=x,y,z)$
satisfying the relation
\begin{align}
    M^{\mathrm{s}}_i(\bq,\Omega) = \chi^{\mathrm{so}}_{ij}(\bq,\Omega) B^{\mathrm{o}}_j(\bq,\Omega) \label{eq:chi-so-definition}
\end{align}
    between the spin magnetization $\boldsymbol{M}^{\mathrm{s}}$ defined in Eq.~(\ref{eq:spin-magnetization})
    and the orbital magnetic field $\bB^{\mathrm{o}}$,
where $\bq$ and $\Omega$ are the wave number (momentum) and the frequency (energy) of the applied magnetic field $\bB^{\mathrm{o}}$.
{As long as the spin polarization is well defined,}
$\chi^{\mathrm{so}}_{ij}$ is uniquely defined
{in both the relativistic and nonrelativistic regimes.}
Below we derive the response function to the orbital magnetic field $\bB^{\mathrm{o}}$
by the perturbative expansion with respect to the vector potential $\bA$,
in a way similar to the perturbative derivation process of
the orbital-orbital susceptibility \cite{Fukuyama_1970,Fukuyama_1971,Koshino_2007}.
% the CME \cite{Kharzeev_2009,Chang_2015,Ma_2015}.

We start with the {translationally invariant} system
described by the momentum-space Hamiltonian
\begin{align}
    \mathcal{H}_0 = \sum_{\bk} \psi^\dag(\bk) H(\bk) \psi(\bk),
\end{align}
with the fermionic field operator $\psi(\bk)$
and the {kernel matrix of Hamiltonian} $H(\bk)$ acting on the components of the field operator.
This assumption applies to both continuum with continuous translational symmetry
and crystals with discrete translational symmetries.
    By diagonalizing the matrix $H(\bk)$,
    we obtain the energy-momentum dispersion $\epsilon_a(\bk)$,
    corresponding to the band dispersion in crystals,
    and the eigenstate $|u_a(\bk)\rangle$,
    which are related by
\begin{align}
    H(\bk) |u_a(\bk)\rangle = \epsilon_a(\bk) |u_a(\bk)\rangle
\end{align}
    with $a$ the label for the eigenstate.
% {We assume that the Fermi energy $\mu$ for the fermions is incorporated in the Hamiltonian $H(\bk)$.}

\begin{figure}[tbp]
    \includegraphics[width=8.4cm]{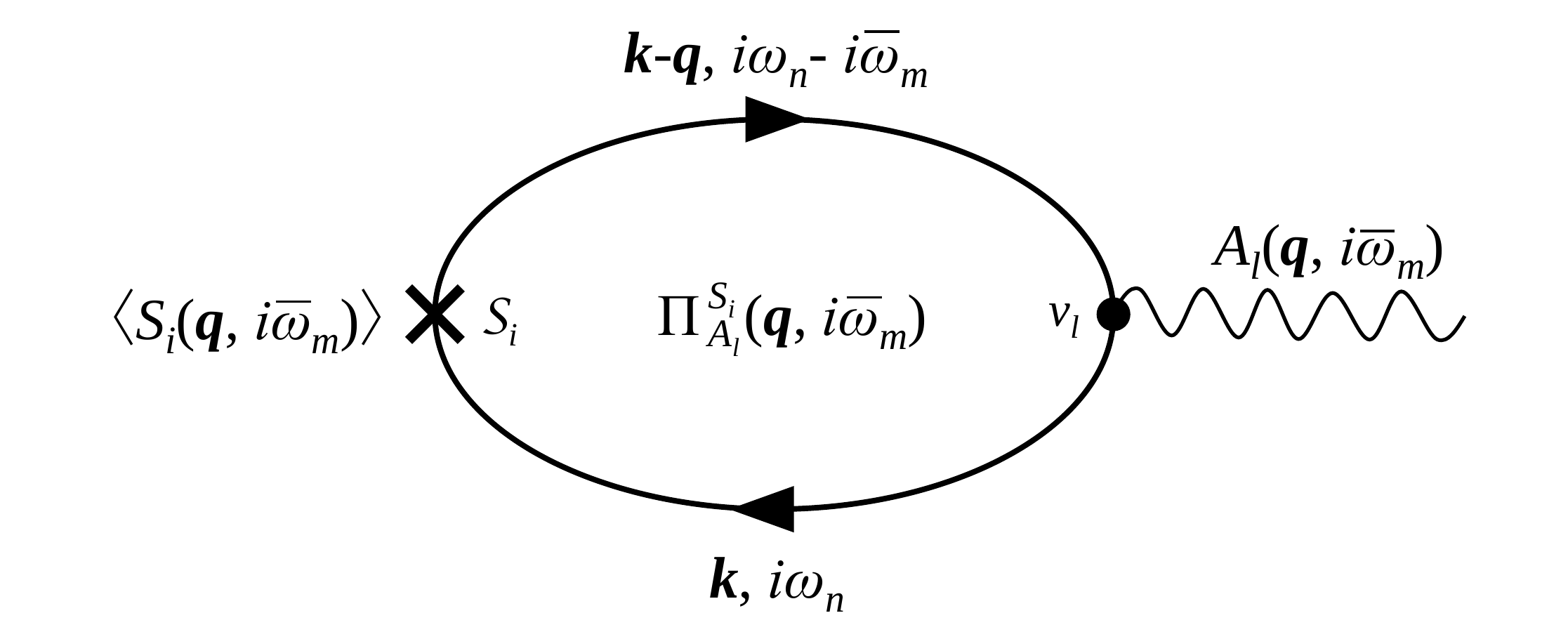}
    \caption{The loop diagram corresponding to the response function $\Pi^{S_i}_{A_l}$ given by Eqs.~(\ref{eq:response-Pi-}) and (\ref{eq:perturbed-spin}),
    with the Matsubara formalism.
    The solid lines represent the fermion propagators,
    and the wavy line corresponds to the external field.}
    \label{fig:loop-diagram}
\end{figure}    

We are here interested in the expectation value of the spin polarization $\langle \boldsymbol{S}(\br,t) \rangle$ under the vector potential $\bA(\br,t)$.
The local spin operator $\boldsymbol{S}(\br,t)$ is defined with the fermionic field operators $(\psi^\dag,\psi)$ as
\begin{align}
    \boldsymbol{S}(\br,t) &= \psi^\dag(\br,t) \bSpin \psi(\br,t),
\end{align}
where $\bSpin$ is the matrix acting on the spin subspace of the fermionic fields,
usually related to the Pauli matrices $\Spin_i = \sigma_i/2$.
As the linear response of spin polarization $\langle S_{i=x,y,z} \rangle$ to the vector potential $A_{l=x,y,z}$,
we focus on the response function $\Pi^{S_i}_{A_l}$ defined by
\begin{align}
    \langle S_i(\br,t) \rangle = \int d\br' dt' \ \Pi^{S_i}_{A_l}(\br-\br',t-t') A_l(\br',t'),
\end{align}
or its Fourier transform
\begin{align}
    \langle S_i (\bq,\Omega) \rangle = \Pi^{S_i}_{A_l}(\bq,\Omega) A_l(\bq,\Omega) \label{eq:response-Pi-}
\end{align}
at arbitrary frequency $\Omega$ and momentum $\bq$.

% in the linear response of the spin polarization against the vector potential,
% \begin{align}
%     \langle S_i (\bq,\Omega) \rangle = \Pi^{S_i}_{A_l}(\bq,\Omega) A_l(\bq,\Omega),
% \end{align}
% % !! definition with Fourier transformation.
% We assume here that the spin polarization operator $S_i$
% appearing in Eq.~(\ref{eq:spin-magnetization})
% is given with the fermionic field operators $(\psi^\dag,\psi)$ and the matrix $\Spin_i$ as
% \begin{align}
%     S_i = \psi^\dag \Spin_i \psi.
% \end{align}

In order to evaluate the response function $\Pi^{S_i}_{A_l}(\bq,\Omega)$,
we first note that the coupling to the vector potential with an arbitrary momentum $\bA(\bq)$ is given as the perturbation term
\begin{align}
    \delta H^{\mathrm{o}}(\bk,\bk') &= \frac{e}{2} \left[ \bv(\bk) + \bv(\bk') \right] \cdot \bA(\bk-\bk'), \label{eq:perturbation-orbital}
\end{align}
% Generally, the coupling to the vector potential with an arbitrary momentum,
% $\bA(\bq) = \int d \boldsymbol{r} \; e^{i \bq \cdot \br} \bA(\br)$,
% is given by the perturbation term
% \begin{align}
%     \delta H^{\mathrm{o}}(\bk,\bk') &= -\frac{e}{2} \left[ \bv(\bk) + \bv(\bk') \right] \cdot \bA(\bk-\bk'), \label{eq:perturbation-orbital}
% \end{align}
with the velocity matrix $\bv(\bk) = \partial H(\bk) / \partial \bk$ [see Eq.~(\ref{eq:mag-orbital})].
Based on this coupling,
the response of the spin polarization $\langle S_i \rangle$ is given with the Matsubara formalism at one loop, {as shown in Fig.~\ref{fig:loop-diagram},}
\begin{align}
    & \langle S_i(\bq,i\bar{\omega}_m)\rangle = -\frac{e A_l(\bq,i\bar{\omega}_m)}{2\beta V} \sum_{i\omega_n,\bk} \Tr \bigl\{ \Spin_i G(\bk,i\omega_n) \label{eq:perturbed-spin} \\
    & \hspace{1.2cm} \times \left[v_l (\bk)+v_l(\bk-\bq)\right] G(\bk-\bq,i\omega_n-i\bar{\omega}_m) \bigr\}, \nonumber
\end{align}
using the unperturbed Green's function $G(i\omega_n,\bk) = [i\omega_n +\mu - H(\bk)]^{-1}$.
Here $\beta = 1/T$ is the inverse temperature,
$V$ is the volume of the system,
{$\mu$ is the chemical potential (Fermi energy) of the fermions,}
and $\bar{\omega}_m$ and $\omega_n$ are bosonic and fermionic Matsubara frequencies, respectively.
We do not consider vertex correction to the velocity vertex,
as long as we omit interaction among fermions or impurity scattering.
% Since we are omitting
% interaction among fermions or impurity scattering,
% our analysis is free from vertex corrections.
By evaluating the Matsubara sum over $i\omega_n$ and performing the {analytical continuation} $i\bar{\omega}_m \rightarrow \Omega + i0$
{(see Appendix \ref{sec:derivation} for detailed derivation process)},
we obtain
\begin{align}
    \Pi^{S_i}_{A_l}(\bq,\Omega) &= -\frac{e}{V}\sum_{\bk} \sum_{ab} \FF_{ab}(\bk,\bq,\Omega) \MM_{ab}^{il}(\bk,\bq). \label{eq:response-Pi}
\end{align}
The factor
\begin{align}
    \MM_{ab}^{il}(\bk,\bq) &= \tfrac{1}{2} \langle u_b(\bk-\bq) | \mathcal{S}_i |  u_a(\bk) \rangle  \label{eq:response-M} \\
    & \hspace{0.5cm} \times \langle u_a(\bk) | v_l (\bk)+v_l(\bk-\bq) | u_b(\bk-\bq)\rangle \nonumber
\end{align}
measures the correlation between spin and orbital degrees of freedom
mediated by the single-particle states $|  u_a(\bk) \rangle$ and $| u_b(\bk-\bq)\rangle$,
and the factor
\begin{align}
    \FF_{ab}(\bk,\bq,\Omega) &= \frac{f(\epsilon_a(\bk)-\mu) - f(\epsilon_b(\bk-\bq)-\mu)}{\epsilon_a(\bk)-\epsilon_b(\bk-\bq) -\Omega -i0} \label{eq:response-F}
\end{align}
specifies the spectral weight from the above two states, {with $f$ the Fermi distribution function.}
% \begin{align}
%     \Pi^{S_i}_{A_l}(\bq,\Omega) &= -\frac{e}{V}\sum_{\bk} \sum_{a,b} \FF_{ab}(\bk,\bq,\Omega) \MM_{ab}^{il}(\bk,\bq) \label{eq:response-Pi} \\
%     % description about each factors...
%     \FF_{ab}(\bk,\bq,\Omega) &= \frac{f(\epsilon_a(\bk)) - f(\epsilon_b(\bk-\bq))}{\epsilon_a(\bk)-\epsilon_b(\bk-\bq) -\Omega -i0} \label{eq:response-F} \\
%     \MM_{ab}^{il}(\bk,\bq) &= \tfrac{1}{2} \langle u_b(\bk-\bq) | \mathcal{S}_i |  u_a(\bk) \rangle  \label{eq:response-M} \\
%     & \hspace{0.5cm} \times \langle u_a(\bk) | v_l (\bk)+v_l(\bk-\bq) | u_b(\bk-\bq)\rangle. \nonumber
% \end{align}

Since the orbital magnetic field $\bB^{\mathrm{o}}$ and the vector potential $\bA$ are related by $\bB^{\mathrm{o}} = \bnab \times \bA$,
or
\begin{align}
    B^{\mathrm{o}}_j(\bq,\Omega) = i \epsilon_{jlh} q_h A_l(\bq,\Omega)
\end{align}
in momentum space,
the susceptibility tensor $\chi^{\mathrm{so}}_{ij}(\bq,\Omega)$
{defined by Eq.~(\ref{eq:chi-so-definition})
can be derived from $\Pi^{\Spin_i}_{A_l}(\bq,\Omega)$ satisfying Eq.~(\ref{eq:response-Pi-})},
\begin{align}
    \chi^{\mathrm{so}}_{ij}(\bq,\Omega) &= \frac{i}{2}\gamma \epsilon_{jlh} \frac{\partial \Pi^{S_i}_{A_l}(\bq,\Omega)}{\partial q_h}. \label{eq:response-chi-so}
\end{align}
% Here we have used Eqs.~(\ref{eq:spin-magnetization}) and (\ref{eq:chi-so-definition}).
Owing to the antisymmetrization by $\epsilon_{jlh}$,
$\chi^{\mathrm{so}}_{ij}(\bq,\Omega)$ becomes gauge independent.

\subsection{Static and dynamical susceptibilities} \label{sec:static-dynamical}

We are mostly interested in the behavior of $\chi^{\mathrm{so}}_{ij}(\bq,\Omega)$
in the low-frequency $(\Omega \rightarrow 0)$ and long-wavelength $(\bq \rightarrow 0)$ limits.
% On evaluating response functions in these {two} limits,
% we should note that, depending on the stage at which the low-frequency limit is evaluated throughout the calculation process,
% it will give different {physical properties} as follows
{
    There are two ways in taking the low-frequency limit,
    which correspond to different physical situations as follows
}
\cite{Luttinger_1964,Mahan,Shastry_2009,Peterson_2010}:
\begin{itemize}
    % equilibrium limit --> static limit
    \item \textit{Static limit} ---
    If one takes $\Omega = 0$ (or $i\bar{\omega}_m=0$) from the beginning,
    the response function gives the behavior of the system in a thermal equilibrium
    that is reconstructed under the external field, {such as the Landau levels under $\bB^{\mathrm{o}}$}.
    This picture is valid if the system thermalizes to the new equilibrium
    as quickly as the external field is applied,
    corresponding to the case $\tau^{-1} \gtrsim \Omega$, where $\tau$ is the relaxation time
    that phenomenologically characterizes the timescale of thermalization process.
    \item \textit{Dynamical limit} ---
    If one keeps $\Omega \neq 0$ at first step
    and takes the limit $\Omega \rightarrow 0$ after evaluating $\bq \rightarrow 0$,
    the response function gives the response arising from the nonequilibrium modulation of the particle distribution,
    driven by the introduction of the external field.
    This picture is valid if the thermalization process is slow enough
    so that the distribution function of particles cannot follow the applied external field,
    corresponding to the case $\tau^{-1} \lesssim \Omega$.
\end{itemize}
In addition to the above conditions,
in order to apply the low-frequency limit $\Omega \rightarrow 0$,
the frequency $\Omega$ should be smaller than the energy scale of band splitting (level repulsion), such as the bandgap.
In the presence of interband hybridization
{(characterized by the parameter $h$ in Sec.~\ref{sec:model}),}
it leads to level repulsion at the band crossing point,
providing one characteristic energy scale for this condition.

While the static limit is mainly considered for the SO crossed susceptibility in previous work \cite{Tserkovnyak_2015,Koshino_2016,Nakai_2016,Suzuura_2016,Ando_2017,Ominato_2019,Aftab_2020,Ozaki_2021},
the dynamical limit is important as well,
since it {is related to experimental measurements} with a magnetic field applied in a short time scale, such as a pulse magnetic field.
Especially, in relativistic heavy-ion collision processes,
the magnetic field is generated soon after the two nuclei collide peripherally,
which is more likely to be described by the dynamical limit.
We therefore consider the SO crossed susceptibilities in both the static and dynamical limits,
which we distinguish by $\chi_{ij}^{\mathrm{so(sta)}}$ and $\chi_{ij}^{\mathrm{so(dyn)}}$,
and compare them in the following discussions.

A difference between the static and dynamical limits emerges in the limiting behavior of the weight factor $\FF_{ab}(\bk,\bq,\Omega)$ given in Eq.~(\ref{eq:response-F}):
\begin{itemize}
    \item \textit{Interband effect} ---
    If two bands $a$ and $b$ are different ($\epsilon_a(\bk) \neq \epsilon_b(\bk)$, or $a \not\equiv b$ for simplicity of notation),
    both the static and dynamical limits give the same factor
    \begin{align}
        \FF_{ab}(\bk,\bq,\Omega)|_{a \not \equiv b} \rightarrow \frac{f(\epsilon_a(\bk)-\mu) - f(\epsilon_b(\bk)-\mu)}{\epsilon_a(\bk) - \epsilon_b(\bk)}.
    \end{align}
    \item \textit{Intraband effect} ---
    If $a$ and $b$ correspond to a same band or degenerate bands ($\epsilon_a(\bk) = \epsilon_b(\bk)$, or $a \equiv b$),
    there arises a difference between the static and dynamical limits.
    By taking $\Omega \rightarrow 0$ first, the static limit gives
    \begin{align}
        \FF_{ab}(\bk,\bq,\Omega)|_{a \equiv b} \rightarrow f'(\epsilon_a(\bk)-\mu),
    \end{align}
    with $f'(\epsilon) = \partial f / \partial \epsilon$,
    since the numerator and the denominator {in $\FF_{ab}$} simultaneously approach zero under $\bq \rightarrow 0$.
    On the other hand, in the dynamical limit, with $\bq \rightarrow 0$ taken first,
    only the numerator approaches zero and this factor vanishes.
\end{itemize}
% On the basis of the semiclassical wave-packet picture,
% the interband and intraband effects are related to the correction in the wave packet constructed from the Bloch states,
% up to the first order in the magnetic field \cite{Gao_2015,Gao_2014}.
The difference in the limiting behavior of the intraband effect results in the difference in the susceptibility,
as we demonstrate in the following discussions.

\subsection{Identification with geometric quantities} \label{sec:geometric-formula}
The SO crossed susceptibility {obtained by} Eq.~(\ref{eq:response-chi-so})
can be further evaluated by expanding the energies and the eigenfunctions by $\bq$ up to its first order.
The $\bq$-expansion
{yields}
the $\bk$-space gradient of the energy $\bnab_{k} \epsilon_a(\bk) = \bv_a(\bk)$, namely the group velocity,
and the gradient of the eigenfunction $|\bnab_{k} u_a(\bk)\rangle$.
In order to rearrange the obtained terms,
it is instructive to introduce the multiband expressions of the geometrical quantities characterizing
the $\bk$-space structure of the wave functions \cite{Chang_1995,Chang_1996,Sundaram_1999,Xiao_2006,Chang_2008,Xiao_2010}.
(Note that these multiband expressions are introduced to simplify the obtained formulas,
and hence are rigorously different from the precise {multiband definitions introduced in Ref.~\onlinecite{Chang_2008}}.)
The physical meanings of the geometric quantities are given in terms of the wave-packet picture \cite{Hagedorn_1980},
where a wave packet localized in real space and momentum space is constructed
as linear combination of the momentum-space wave functions $|u_a(\bk)\rangle$.
(For simplicity of notations,
we do not explicitly {denote the argument $\bk$} below.)

We here introduce the Berry connection $\bA_{ab}$, the orbital magnetic moment $\bmag_{ab}$,
the Berry curvature $\bOm_{ab}$, and the spin Berry curvature $\bOm^{(\Spin_i)}_{ab}$.
Below we give their definitions
and their physical meanings on the basis of the wave-packet picture.
The Berry connection
\begin{align}
    \bA_{ab} &= i \langle u_a | \bnab_k u_b \rangle,
\end{align}
namely the matrix element of the position operator $\br = i\bnab_k$,
is related to the shift of the wave-packet center in real space due to the quantum interference.
The orbital magnetic moment {is defined as}
\begin{align}
    \bmag_{ab} &= \frac{ie}{2} \langle \bnab_k u_a | \times ( \bar{\epsilon}_{ab} - H ) | \bnab_k u_{b} \rangle,
\end{align}
with $\bar{\epsilon}_{ab} = (\epsilon_a + \epsilon_b)/2$.
(The cross product acts on the Cartesian components arising from the momentum gradient $\bnab_k = \sum_{j=x,y,z} \boldsymbol{e}_j \partial_{k_j}$,
where $\boldsymbol{e}_{x,y,z}$ are the unit vectors in the Cartesian coordinate.)
$\bmag_{ab}$ is related to the orbital angular momentum intrinsic to the wave packet,
which arises from geometrical structure of the wave functions.
It is in analogy with the ``self-rotation'' of a classical rigid body,
and is distinct from motion of the wave-packet center \cite{Chang_1996,Sundaram_1999,Xiao_2006,Xiao_2010}.
The Berry curvature
\begin{align}
    \bOm_{ab} &= i \langle \bnab_k u_a | \times | \bnab_k u_b \rangle
\end{align}
and the ``spin Berry curvature''
\begin{align}
    \bOm^{(\Spin_i)}_{ab} &= i \langle \bnab_k u_a | \times \Spin_i | \bnab_k u_b \rangle
\end{align}
roughly correspond to the circulating current and spin current, respectively,
arising from the geometrical structure of the wave functions.
Since all these effects couple to the vector potential or the orbital magnetic field in real space,
these geometric quantities appear in the response functions to the orbital magnetic field.

Using the expressions of the geometric quantities introduced above,
we can classify the SO crossed susceptibility into three terms,
\begin{align}
    \chi_{ij}^{\mathrm{so}} &= \chi_{ij}^{(A)} + \chi_{ij}^{(m)} + \chi_{ij}^{(\Omega)},
\end{align}
where the first term picks up the contribution from the Berry connection,
the second term from the orbital magnetic moment,
and the third term from the Berry curvature and the spin Berry curvature.
{(The detailed calculation process is shown in Appendix \ref{sec:geometric}.)}
Here we introduce the shorthand notations for
%  the group velocity $\bv_a = \bnab_k \epsilon_a$,
the spin matrix element
$\Spin_{ab}^i = \langle u_a | \Spin_i | u_b \rangle$,
the Fermi distribution function $f_a = f(\epsilon_a(\bk)-\mu)$,
and the weight factor
\begin{align}
    F_{ab} = 
    \begin{cases}
        f'_a & (a \equiv b) \\
        \displaystyle \frac{f_a-f_b}{\epsilon_a-\epsilon_b} & (a \not\equiv b)
    \end{cases}
\end{align}
arising from $\FF_{ab}$ in Eq.~(\ref{eq:response-F}).
With these notations, the Berry connection term is given as
\begin{align}
    \chi_{ij}^{\mathrm{so(sta:}A)} &= -\frac{e \gamma}{V} \sum_{\bk}  \sum_{a \not \equiv b}(f'_a-F_{ab}) \re[(\bv_a \times \bA_{ab})^j \Spin_{ba}^i] \\
    \chi_{ij}^{\mathrm{so(dyn:}A)} &= -\frac{e \gamma}{V} \sum_{\bk}  \sum_{a \not\equiv b}(\tfrac{1}{2}f'_a-F_{ab}) \re[(\bv_a \times \bA_{ab})^j \Spin_{ba}^i],
\end{align}
and the orbital magnetic moment term as
\begin{align}
    \chi_{ij}^{\mathrm{so(sta:}m)} &= \frac{\gamma}{V} \sum_{\bk} \sum_{a b} F_{ab} \re\left[ m_{ab}^j \Spin_{ba}^i\right] \label{eq:chi-sta-m} \\
    \chi_{ij}^{\mathrm{so(dyn:}m)} &= \frac{\gamma}{V} \sum_{\bk} \sum_{a \not\equiv b} F_{ab} \re\left[m_{ab}^j \Spin_{ba}^i\right], \label{eq:chi-dyn-m}
\end{align}
in the static and dynamical limits, respectively.
% These terms contain both intraband $(a \equiv b)$ and interband $(a \not\equiv b)$ contributions in the static limit,
% whereas they contain only the interband contributions in the dynamical limit,
% which can be traced back to the weight factor $\FF$ in Eq.~(\ref{eq:response-F}).
The Berry curvature term
\begin{align}
    \chi_{ij}^{\mathrm{so}(\Omega)} &= -\frac{e \gamma}{2V} \sum_{\bk} \biggl[ \sum_{ab} f_a \re\left(\Omega_{ab}^j \Spin_{ba}^i\right) +\sum_a f_a \Omega_{aa}^{(\Spin_i)j} \biggr],
\end{align}
arising from the Berry curvature and the spin Berry curvature,
takes the same form for the static and dynamical limits,
since this term is originally proportional to $f_a-f_b$
and the intraband effect $a \equiv b$ gives no contribution to this term.

The static SO susceptibility,
namely the response of the spin magnetization to the orbital magnetic field {in equilibrium},
is equivalent to its counterpart in terms of the Onsager's reciprocity theorem \cite{Onsager}:
the response of the orbital magnetization in equilibrium \cite{Xiao_2010,Thonhauser_2005,Ceresoli_2006,Shi_2007}
\begin{align}
    \boldsymbol{M}^{\mathrm{o}} &= -\frac{ie}{2V} \sum_{a, \bk} f_a \langle \bnab_k u_a | \times (\epsilon_a + H -2\mu) | \bnab_k u_a \rangle \label{eq:M-orbital}
\end{align}
to the spin magnetic field (Zeeman splitting)
\begin{align}
    \delta \mathcal{H}^{\mathrm{s}} = \gamma \int d\br \boldsymbol{B}^{\mathrm{s}} \cdot \boldsymbol{S},
\end{align}
consistently reproduces the static susceptibility obtained above [see Eq.~(\ref{eq:Onsager})].
Since the formula Eq.~(\ref{eq:M-orbital}) is valid only in equilibrium,
we cannot rederive the dynamical SO crossed susceptibility,
    which is based on the nonequilibrium distribution disturbed by the magnetic field,
from this reciprocity.

% The principal difference between the equilibrium and dynamical susceptibilities is
% the presence or absence of the intraband process $a \equiv b$,
% as mentioned in the previous subsection.
% While the Berry connection and magnetic moments terms in equilibrium have contributions
% from both the intraband $(a \equiv b)$ and interband $(a \not\equiv b)$ processes,
% those in the dynamical limit are dominated by the interband processes.
% The intraband contribution is influenced by the structure of the Fermi surface $f'_a$,
% whereas the interband contribution is accompanied with the factor $F_{ab} = (f_a - f_b)/(\epsilon_a - \epsilon_b)$;
% their difference arises crucially in the Fermi-level dependence of $\chi^{\mathrm{so}}$,
% which shall be demonstrated in the following discussions.

The Berry-curvature term, arising from all the occupied states in the Fermi sea,
contributes to the static and dynamical susceptibilities in the same manner.
Since it counts up the entire contribution from the Fermi sea,
the susceptibility depends on the momentum-space cutoff,
which corresponds to the structure of the Brillouin zone in crystals.
In order to extract the universal behavior arising from the relativistic dispersion and the band hybridization in later sections,
we do not concentrate on the value of $\chi_{ij}^{\mathrm{so}}$ itself,
but discuss its dependence on the Fermi level $\mu$ throughout this article.
We evaluate its deviation from the value at $\mu =0$,
\begin{align}
    \Delta{\chi}^{\mathrm{so}}_{ij}(\mu) = {\chi}^{\mathrm{so}}_{ij}(\mu) - {\chi}^{\mathrm{so}}_{ij}(\mu=0),
\end{align}
which is the quantity considered in Ref.~\onlinecite{Koshino_2016} for Dirac and Weyl fermions.

\section{Single Weyl node} \label{sec:weyl}
Before going on to detailed analysis with interband hybridization effect,
let us check how the above formula works
by taking a single Weyl node as a simplest test case,
which is in parallel with the analyses in Refs.~\onlinecite{Nomura_2015,Koshino_2016}.
We shall see that, due to spin-momentum locking,
the SO crossed response of Dirac and Weyl fermions
is closely related to the chiral magnetic effect (CME) and the chiral separation effect (CSE),
which are {the response phenomena with respect to the orbital magnetic field},
well studied in the context of relativistic field theory \cite{Vilenkin_1980,Nielsen_1983,Kharzeev_2006,Kharzeev_2007,Kharzeev_2008,Fukushima_2008}.

In lattice systems,
the Nielsen--Ninomiya theorem \cite{Nielsen_1981_1,Nielsen_1981_2} requires that Weyl nodes with opposite chiralities should appear in pairs.
We can still rely on the single Weyl-node picture,
as long as we neglect the contribution from $\bk$ away from the Weyl points and extract the $\mu$-dependence.
This picture is valid if the Weyl points are well separated in momentum space.
At momenta $\bk$ away from the Weyl points,
the corresponding band energy $\epsilon_a(\bk)$ should be located far above or below the Fermi level $\mu$,
so that the region around the Weyl points will give the dominant contribution to the response phenomenon.
Under such conditions, we can treat the quasiparticle excitations around each Weyl point separately.

\subsection{SO crossed susceptibility}
If we assume spherical symmetry around the Weyl points,
we can use the momentum-space Hamiltonian as a $2\times 2$-matrix,
\begin{align}
    H (\bk) &= \eta \vf \bk \cdot \bsig,
\end{align}
where the momentum $\bk$ is defined as the relative momentum from the Weyl point,
with the spherical coordinate $\bk = k(\sin\theta\cos\phi, \sin\theta\sin\phi, \cos\theta)$.
This matrix acts on the spin-$1/2$ space with spin-up and down states,
{where the Pauli matrix $\bsig$ corresponds to the spin operator $\bSpin$ by $\bSpin = \bsig/2$}.
The chirality (right/left) for each Weyl node is identified by $\eta = \pm$,
and $\vf$ denotes the Fermi velocity around the Weyl point.
This Hamiltonian yields the conventional linear dispersion,
with the positive-energy branch $\epsilon(\bk) = \vf |\bk|$ and negative-energy branch $\epsilon(\bk) = -\vf |\bk|$,
corresponding to the eigenfunctions
\begin{align}
    |u_+(\bk)\rangle =
    \begin{pmatrix}
        e^{-i\phi/2} \cos\frac{\theta}{2} \\ e^{i\phi/2} \sin\frac{\theta}{2}
    \end{pmatrix}
    ,
    |u_-(\bk)\rangle =
    \begin{pmatrix}
        e^{-i\phi/2} \sin\frac{\theta}{2} \\ -e^{i\phi/2} \cos\frac{\theta}{2}
    \end{pmatrix}
    .
\end{align}
For the chirality $\eta = +$,
the positive-energy branch corresponds to $|u_+\rangle$ and the negative-energy branch to $|u_-\rangle$,
and vice versa for $\eta = -$.
Taking the eigenstates $|u_\pm\rangle$ as the basis,
the intraband and interband geometrical quantities within the single Weyl node are given as
\begin{align}
    & \bA_{\pm,\pm} = \frac{\pm 1}{2k} \cot\theta \ep, \quad
    \bA_{\pm,\mp} = \frac{1}{2k} (\pm i\et + \ep), \\
    & \bmag_{\pm,\pm} = -\frac{\eta e \vf}{2k} \ek, \quad
    \bmag_{\pm,\mp} = 0, \label{eq:Dirac-m} \\
    & \bOm_{\pm,\pm} = \frac{\mp 1}{2k^2} \ek, \quad
    \bOm_{\pm,\mp} = \frac{1}{2k^2} \cot\theta \ek, \\
    & \bOm^{(\Spin_i)}_{\pm,\pm} = \frac{1}{4k^2}(e_k^i -e_\theta^i \cot\theta) \ek,
\end{align}
with the unit vectors
$\ek = \bk / |\bk|$, $\ep = \boldsymbol{e}_z \times \ek/\sin\theta$, $\et = \ep \times \ek$.
The matrix elements of the spin operator $\bSpin = \bsig/2$ read
\begin{align}
    \bSpin_{\pm,\pm} = \pm \frac{1}{2} \ek, \ 
    \bSpin_{\mp,\pm} = \mp \frac{1}{2} ( i \ep - \et). \label{eq:Dirac-spin}
\end{align}

Using the above geometrical quantities and matrix elements,
the SO crossed susceptibility tensor for a single Weyl node,
at Fermi level $\mu$,
can be straightforwardly obtained,
\begin{align}
    \Delta {\chi}^{\mathrm{so(sta)}}_{ij}(\mu) &= \frac{e\gamma\mu}{8\pi^2 \vf} \delta_{ij}, \label{eq:chi-eq-Dirac} \\
    \Delta {\chi}^{\mathrm{so(dyn)}}_{ij}(\mu) &= \frac{e\gamma\mu }{24\pi^2 \vf} \delta_{ij}, \label{eq:chi-dyn-Dirac}
\end{align}
at zero temperature.
% where $\mathbf{I}$ is the identity tensor $[\mathbf{I}]_{ij} = \delta_{ij}$.
The static susceptibility, given by Eq.~(\ref{eq:chi-eq-Dirac}),
correctly reproduces the result in Ref.~\onlinecite{Koshino_2016} obtained by explicitly counting up the contributions from the Landau levels under the magnetic field.
On the other hand, the magnitude of the dynamical susceptibility given by Eq.~(\ref{eq:chi-dyn-Dirac}) is one third that of the static susceptibility,
which has not been explicitly mentioned in the context of the SO crossed susceptibility.
Such a difference between static and dynamical limits is also seen in the CME and the CSE,
% and the response of ``chiral'' current, namely the chiral separation effect (CSE),
which we shall discuss in detail below.
The $\mu$-dependence in the SO crossed susceptibility of Weyl fermions
implies that,
under an orbital magnetic field $\bB^{\mathrm{o}}$,
spin polarization of the Weyl fermions can be induced by varying the chemical potential $\mu$.
This electron spin polarization will exert a spin torque on magnetization
if the system has a ferromagnetic order,
which is proposed as the charge- or voltage-induced torque 
in the context of magnetic Weyl semimetals \cite{Nomura_2015}.

\subsection{Chiral magnetic/separation effects}
The SO crossed susceptibility of Dirac and Weyl fermions is {closely related to the CME,
namely the current response against an orbital magnetic field} \cite{Vilenkin_1980,Nielsen_1983,Kharzeev_2006,Kharzeev_2007,Kharzeev_2008,Fukushima_2008}.
For a single Weyl node with chirality $\eta$,
the current operator $\boldsymbol{j}$ and the spin operator $\boldsymbol{S}$ are related as
\begin{align}
    \boldsymbol{j} &= -\eta e \vf \psi^\dag \bsig \psi = -2\eta e \vf \boldsymbol{S}.
\end{align}
Therefore, when the spin polarization
\begin{align}
    \langle \boldsymbol{S} \rangle_\eta^{\mathrm{(sta)}} = -\frac{e\mu}{8\pi^2 \vf} \bB^{\mathrm{o}} , \quad
    \langle \boldsymbol{S} \rangle_\eta^{\mathrm{(dyn)}} = -\frac{e\mu}{24\pi^2 \vf} \bB^{\mathrm{o}}
\end{align}
is induced by the magnetic field $\bB^{\mathrm{o}}$,
the chirality-dependent current
\begin{align}
    \langle \boldsymbol{j} \rangle_\eta^{\mathrm{(sta)}} = \frac{\eta e^2\mu}{4\pi^2} \bB^{\mathrm{o}} , \quad
    \langle \boldsymbol{j} \rangle_\eta^{\mathrm{(dyn)}} = \frac{\eta e^2\mu}{12\pi^2} \bB^{\mathrm{o}}
\end{align}
is induced accordingly.
For a pair of Weyl nodes, or a single Dirac node,
the net current vanishes once it is summed over the chirality $\eta=\pm$.
In case the chemical potentials of the two chiralities $\mu_{\eta=\pm}$ are different,
which is characterized by the chiral chemical potential $\mu_5 = (\mu_+ - \mu_-)/2$,
the net current does not fully cancel.
The current $\langle \boldsymbol{j} \rangle = \langle \boldsymbol{j} \rangle_+ + \langle \boldsymbol{j} \rangle_-$ arises in response to $\bB^{\mathrm{o}}$,
\begin{align}
    \langle \boldsymbol{j} \rangle^{\mathrm{(sta)}} = \frac{e^2\mu_5}{2\pi^2} \bB^{\mathrm{o}} , \quad
    \langle \boldsymbol{j} \rangle^{\mathrm{(dyn)}} = \frac{e^2\mu_5}{6\pi^2} \bB^{\mathrm{o}},
\end{align}
which is consistent with the static and dynamical CME obtained by the field-theoretical approach \cite{Kharzeev_2009,Ma_2015} and the semiclassical approach \cite{Son_2013,Chang_2015,Zhong_2016}.
{We should be careful} that the CME in equilibrium is unrealistic in lattice systems,
once one takes into account all the occupied states below the Fermi level,
including the states away from the Weyl points \cite{Vazifeh_2013,Basar_2014,Landsteiner_2014,Landsteiner_2016}.
On the other hand, the dynamical CME,
which is explicitly referred to as the gyrotropic magnetic effect (GME) \cite{Zhong_2016}
or the natural optical activity \cite{Ma_2015} as well, is still present in {crystals},
arising from the field-induced modulation of the density of states at the Fermi surfaces.

In the absence of $\mu_5$,
the charge current $\langle \boldsymbol{j} \rangle$ vanishes in total,
whereas the chiral current $\langle \boldsymbol{j}_5 \rangle = \langle \boldsymbol{j} \rangle_+ - \langle \boldsymbol{j} \rangle_-$,
{corresponding to the currents of right-handed and left-handed fermions flowing in opposite directions}, is present.
The chiral current arises in response to the orbital magnetic field $\bB^{\mathrm{o}}$,
\begin{align}
    \langle \boldsymbol{j}_5 \rangle^{\mathrm{(sta)}} = \frac{e^2\mu}{2\pi^2} \bB^{\mathrm{o}} , \quad
    \langle \boldsymbol{j}_5 \rangle^{\mathrm{(dyn)}} = \frac{e^2\mu}{6\pi^2} \bB^{\mathrm{o}},
\end{align}
which is known as the chiral separation effect (CSE) in the context of the relativistic field theory \cite{Son_2004,Melitski_2005}.
The difference between the static and dynamical limit is present in the CSE as well \cite{Landsteiner_2014_2}.
Note that the chiral current $\langle \boldsymbol{j}_5 \rangle$ is proportional to the net spin polarization $\langle \boldsymbol{S} \rangle = \langle \boldsymbol{S} \rangle_+ + \langle \boldsymbol{S} \rangle_-$,
\begin{align}
    \langle \boldsymbol{j}_5 \rangle = \sum_\eta \eta \langle \boldsymbol{j} \rangle_\eta
    = \sum_\eta \eta (-2\eta e \vf) \langle \boldsymbol{S} \rangle_\eta = -2 e\vf \langle \boldsymbol{S} \rangle,
\end{align}
due to spin-momentum locking.
While the definition of the SO susceptibility is valid as long as the particles have spin degrees of freedom,
the CSE is well defined only if the chirality is defined as a good quantum number,
and hence we can regard the CSE as the typical example of the SO crossed response
arising exclusively for chiral fermions.

\section{Band hybridization effect} \label{sec:model}
Based on the general formula obtained above,
we now discuss the behavior of the SO crossed susceptibility,
in the hybrid system of Dirac and nonrelativistic fermions.
Using a minimal model Hamiltonian including Dirac and nonrelativistic fermions,
\modify{
    we discuss the effect of band hybridization on the SO susceptibility.
}
We evaluate both the static and dynamical susceptibilities $\Delta \chi^{\mathrm{so(sta/dyn)}}$
as functions of the Fermi energy $\mu$,
and discuss how they get modified from those for Dirac and Weyl fermions
mentioned in the previous section.
% We find that the deviation develops at the Fermi level in the vicinity of the band hybridization points,
% {which appears significantly for the dynamical susceptibility rather than the static susceptibility.}
We shall also separate the {induced spin polarization} into that from the Dirac bands and that from the nonrelativistic bands,
to see {the hybridization-induced effect in more detail.}
    We mainly consider their behavior at zero temperature.
% how much spin polarization can be induced {for each sector}.
% {We find that the dynamical spin polarization of the nonrelativistic fermions
% is induced with the sign opposite to that of the Dirac fermions,
% suppressing the dynamical SO crossed susceptibility in total.}

\subsection{Minimal model}
Let us take into account a single species of Dirac fermions and a single species of nonrelativistic fermions, both with spin $1/2$, in three dimensions.
Here the field operators consist of six components {in total}:
the four-component Dirac sector is labeled by chirality (R/L) and spin $(\uparrow/\downarrow)$ in the Weyl representation,
\begin{align}
    \Psi_{\mathrm{Dirac}} = (\psi_{\mathrm{R}\uparrow},\psi_{\mathrm{R}\downarrow},\psi_{\mathrm{L}\uparrow},\psi_{\mathrm{L}\downarrow})^T,
\end{align}
while the two-component nonrelativistic (NR) sector is labeled by spin $(\uparrow/\downarrow)$,
\begin{align}
    \Psi_{\mathrm{NR}} = (\psi_{\mathrm{NR}\uparrow},\psi_{\mathrm{NR}\downarrow})^T.
\end{align}
{We define the model Hamiltonian for each sector as}
\begin{align}
    \Ham_{\mathrm{Dirac}} &= \int d^3\br \ \Psi_{\mathrm{Dirac}}^\dag(\br) ( -i \vf \boldsymbol{\nabla} \cdot \boldsymbol{\alpha}) \Psi_{\mathrm{Dirac}}(\br) \\
    \Ham_{\mathrm{NR}} &= \int d^3\br \ \Psi_{\mathrm{NR}}^\dag(\br) \left[\frac{-\nabla^2}{2m} + \epsilon_0 \right]\Psi_{\mathrm{NR}}(\br).
\end{align}
The $\alpha$-matrices for the Dirac sector are defined with the Weyl representation,
$\boldsymbol{\alpha} = \mathrm{diag}(\bsig, -\bsig)$.
Here we {assume that} the momentum is locked
not to the pseudospin, such as atomic orbital or sublattice degrees of freedom in crystals,
but to the real spin,
so that $\bsig$ acts on the real spin degrees of freedom \cite{Note-pseudospin}.
For simplicity of discussion, we set the Fermi velocity $\vf$ for the Dirac sector
isotropic around the Dirac point $\bk =0$,
and we take the free-particle dispersion for the nonrelativistic sector.
$m$ denotes the effective mass at band bottom,
and $\epsilon_0$ is the energy difference (offset) of the band
from the Dirac point.

We now take into account hybridization between the Dirac and nonrelativistic bands,
and consider its effect on the SO crossed susceptibility.
The hybridization arises if there is a slight violation of crystalline symmetries that protect the Dirac-node structure,
or an interaction between the Dirac and nonrelativistic sectors.
Whereas its detailed structure depends on the microscopic properties,
namely the crystal structure,
angular momenta of the constituent atomic orbitals, etc.,
our main interest is rather conceptual,
to see
the behavior of the SO crossed susceptibility in the vicinity of the band hybridization point.
We therefore set up a simple structure of hybridization,
which satisfies spherical symmetry, conserves spin,
and acts on the right-handed and left-handed components with the same weights.
The hybridization term is then parametrized by a single real value $h$,
with
\begin{align}
    \Ham_{\mathrm{hyb}} &= h \int d^3\br \sum_{s=\uparrow,\downarrow} \left[ \psi_{\mathrm{R},s}^\dag \psi_{\mathrm{NR},s} + \psi_{\mathrm{L},s}^\dag \psi_{\mathrm{NR},s} + \mathrm{H.c.} \right].
\end{align}
With this hybridization term,
the total Hamiltonian $\Ham = \Ham_{\mathrm{Dirac}} + \Ham_{\mathrm{NR}} + \Ham_{\mathrm{hyb}}$ can be written as a $6\times 6$-matrix in momentum-space representation as follows:
\begin{align}
    & \Ham = \sum_{\bk} \Psi^\dag(\bk) H(\bk) \Psi(\bk), \label{eq:H-full} \\
    & H(\bk) = 
    \begin{pmatrix}
        \vf \bk \cdot \bsig & 0 & h \\
        0 & -\vf \bk \cdot \bsig & h \\
        h & h & \frac{k^2}{2m} + \epsilon_0
    \end{pmatrix} , \label{eq:H-matrix}
\end{align}
with the field operator
\begin{align}
    \Psi = (\psi_{\mathrm{R}\uparrow},\psi_{\mathrm{R}\downarrow},\psi_{\mathrm{L}\uparrow},\psi_{\mathrm{L}\downarrow},
    \psi_{\mathrm{NR}\uparrow},\psi_{\mathrm{NR}\downarrow})^T. \label{eq:field-operator}
\end{align}

\begin{figure}[tbp]
    \includegraphics[width=8.4cm]{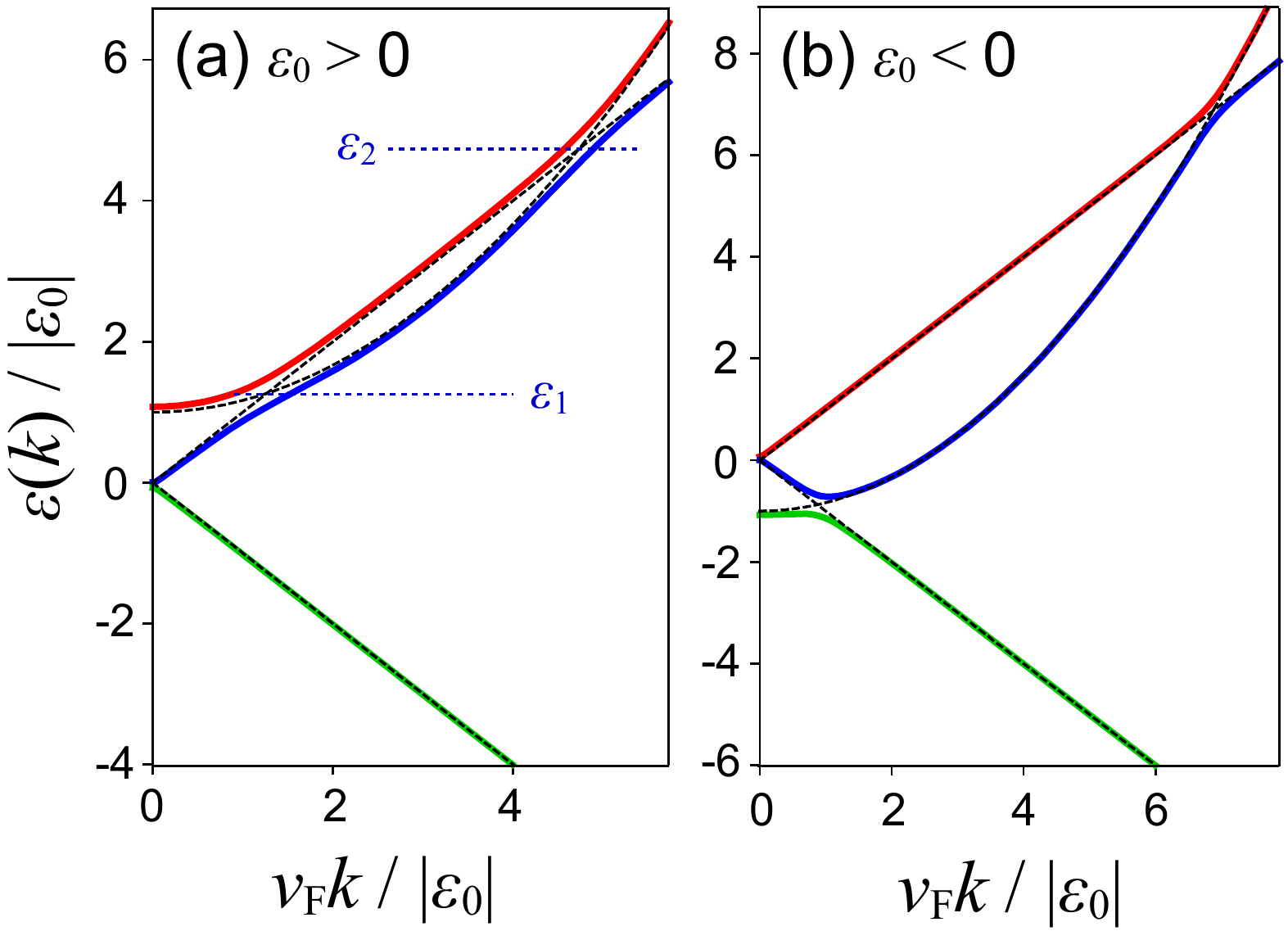}
    \caption{Band structure of the minimal hybridized model defined by Eqs.~(\ref{eq:H-full}) and (\ref{eq:H-matrix}), for the energy offset (a) $\epsilon_0>0$ and (b) $\epsilon_0<0$.
    The parameters are taken as $m = 3|\epsilon_0|$ and $h = 0.2|\epsilon_0|$.
    The dashed lines show the bands without the hybridization $h$.
    $\epsilon_1$ and $\epsilon_2$ in (a) are the energies of the band crossing points in the absence of hybridization,
    which we shall use in later calculations.}
    \label{fig:bands}
\end{figure}     

Since this model Hamiltonian keeps the right-handed and left-handed components in the Dirac sector equivalent,
it yields three bands,
each of which is twofold degenerate and spherically symmetric around $\bk =0$.
Here we note that $H(\bk)$ commutes with the operator $\ek \cdot \bsig$,
corresponding to the helicity of a particle.
Therefore, $H(\bk)$ can be separated into two helicity subspaces characterized by the eigenvalue $\eta = \pm$,
with the $3 \times 3$-matrix
\begin{align}
    H_\eta(k) = 
    \begin{pmatrix}
        \eta \vf k & 0 & h \\
        0 & -\eta \vf k & h \\
        h & h & \frac{k^2}{2m} + \epsilon_0
    \end{pmatrix}
\end{align}
for each subspace.
The typical band structure based on this Hamiltonian is shown by Fig.~\ref{fig:bands}.
In the present model, the quadratic dispersion from the nonrelativistic sector
coexists with the linear dispersion from the Dirac sector.
Therefore,
if the energy offset $\epsilon_0$ is positive,
{as shown in Fig.~\ref{fig:bands}(a),}
the nonrelativistic band intersects the particle (electron) branch of the Dirac band twice,
whose energy levels are labeled as $\epsilon_1$ and $\epsilon_2$ in the following discussions.
At these points, the bands develop anticrossing with the amplitude $h$.
If $\epsilon_0$ is negative,
{as shown in Fig.~\ref{fig:bands}(b),}
the nonrelativistic band intersects the antiparticle (hole) and particle branches of the Dirac band once for each,
yielding a gap at the crossing point with the antiparticle branch.
In the present calculation,
we take $\epsilon _0$ positive,
and investigate the behavior of the SO crossed susceptibility mainly around the hybridization points $\epsilon_{1,2}$.
% In the present calculation,
% we first fix the sign of $\epsilon_0$ and the momenta for the crossing points,
% which we denote $k_1$ and $k_2$, as the input parameters,
% and determine the values of $m$ and $\epsilon_0$ accordingly.

\modify{
    Our model defined here is composed of minimal number of degrees of freedom,
    in order to extract the common feature in the SO susceptibility
    caused solely by the presence or absence of the band hybridization.
    While realistic systems including relativistic fermions generally have richer internal degrees of freedom,
    such as orbital and sublattice degrees of freedom for electrons in solid states
    and color and flavor degrees of freedom for quarks in high-energy physics,
    dependence on such detailed internal structures for each system is beyond our interest in this article.
}

\subsection{Static and Dynamical suceptibilities}

\begin{figure}[tbp]
    \includegraphics[width=8.4cm]{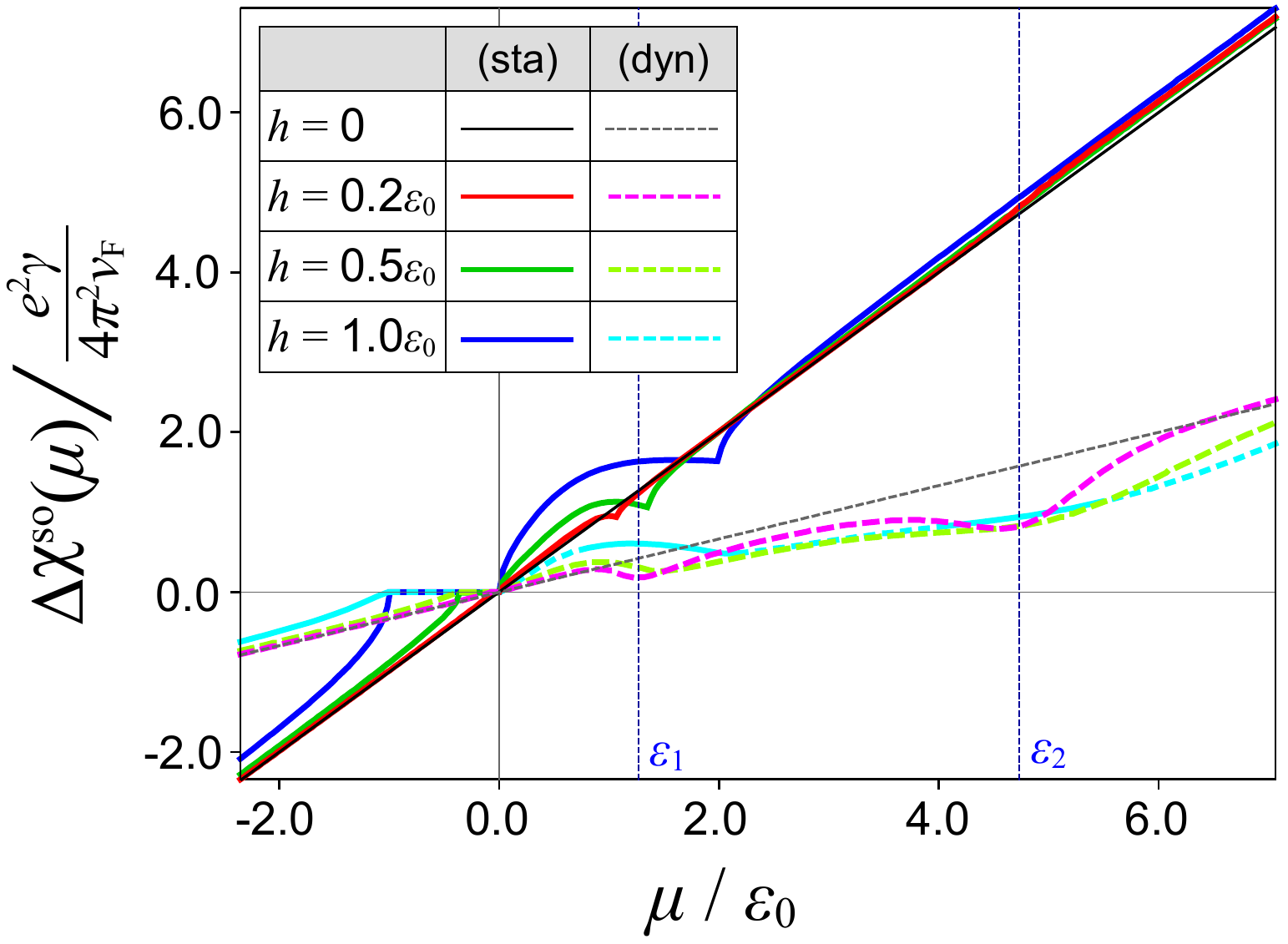}
    \caption{Behavior of the SO crossed susceptibility $\Delta \chi^{\mathrm{so}}(\mu)$, as functions of the Fermi energy $\mu$.
    The solid and dashed lines are the static and dynamical susceptibility, respectively,
    with the hybridization parameter $h$ varied as shown in the inset table.
    The vertical dashed lines $(\epsilon_{1,2})$ correspond to the band hybridization points,
    which are identical to those shown in Fig.~\ref{fig:bands}(a).
    The parameters are taken as $\epsilon_0>0$ and $m = 3\epsilon_0$.}
    \label{fig:hybb-pos-all}
\end{figure}     

With the model Hamiltonian defined in the previous subsection,
we now evaluate the SO crossed susceptibility $\Delta \chi^{\mathrm{so}}(\mu)$ in both static and dynamical limits.
We first consider {the response of} the net spin polarization,
{by taking the spin operator $\bSpin$ as a matrix $\mathrm{diag}(\bsig,\bsig,\bsig)/2$ acting on the 6-component field operator $\Psi$ in Eq.~(\ref{eq:field-operator})}.
Here we fix the band parameters $\epsilon_0 >0$ and $m = 3\epsilon_0$,
and vary the band hybridization parameter $h$ and the Fermi energy $\mu$
to capture typical structures in $\Delta \chi^{\mathrm{so}}(\mu)$,
arising from the band hybridization.
    We evaluate the formula obtained in Sec.~\ref{sec:geometric-formula} numerically,
    based on the band eigenstates of the model Hamiltonian.
The quantities with energy dimensions are rescaled by $\epsilon_0$ in the present calculations.
Since the system is assumed to satisfy spherical symmetry,
the susceptibility tensor possesses only the diagonal part
$\Delta \chi^{\mathrm{so}}_{ij}(\mu) = \Delta \chi^{\mathrm{so}}(\mu) \delta_{ij}$,
which we shall evaluate in the following discussion.

We first compare the static susceptibility $\Delta \chi^{\mathrm{so(sta)}}$
and the dynamical susceptibility $\Delta \chi^{\mathrm{so(dyn)}}$ under the band hybridization,
with
    those estimated with the Dirac fermions without hybridization,
    which we call the ``pure Dirac'' case [Eqs.~(\ref{eq:chi-eq-Dirac}) and (\ref{eq:chi-dyn-Dirac})].
The results are shown in Fig.~\ref{fig:hybb-pos-all} as functions of $\mu$.
In the vicinity of the band hybridization points $\epsilon_{1,2}$,
both the static susceptibility $\Delta \chi^{\mathrm{so(sta)}}$
and the dynamical susceptibility $\Delta \chi^{\mathrm{so(dyn)}}$
deviate from those in the pure Dirac case.
They asymptotically reach {the pure Dirac behavior} at the energies away from $\epsilon_{1,2}$,
since the hybridization effect on the band eigenstates is significant only around these points.

For the static susceptibility $\Delta \chi^{\mathrm{so(eq)}}(\mu)$,
we find {three nonanalytic cusps for each value of $h$.}
{These cusps corresponds} to the band edges,
namely {the minima and maxima of the bands under} the hybridization.
    The origin of such a nonanalytic behavior can be traced back to the density of states,
    which becomes nonanalytic at each band edge.
    We can see that it comes from the intraband part of the magnetic-moment contribution $\chi_{ij}^{\mathrm{so(sta:}m)}$ given by Eq.~(\ref{eq:chi-sta-m}),
    since it is accompanied with the factor $f'(\epsilon_a)$
    that gives the density of states at zero-temperature limit.
% This nonanalyticity stems from the Fermi-surface contribution accompanied with $f'(\epsilon_a)$ in the intraband process,
% {in particular from the magnetic-moment contribution $\chi_{ij}^{\mathrm{so(sta:}m)}$ given by Eq.~(\ref{eq:chi-sta-m})},
% since the density of states at the Fermi energy
% becomes nonanalytic at the band edge.
Nonanalyticity in the static susceptibility is also found for massive Dirac fermions \cite{Koshino_2016},
arising at the edges of the mass gap,
which can also be attributed to the above mechanism.

In contrast,
%  to such a nonanalytic behavior in the equilibrium susceptibility,
the dynamical susceptibility $\Delta \chi^{\mathrm{so(dyn)}}(\mu)$ is obtained as a smooth function in $\mu$,
since it does not contain the intraband Fermi-surface contribution.
Although it still contains the Fermi-surface effect $f'(\epsilon_a)$ in $\chi_{ij}^{\mathrm{so(dyn:}A)}$,
the velocity $\bv_a$ in the same term reaches zero at the band edge,
canceling nonanalyticity from the density of states.

Aside from the nonanalyticity,
we should note that the dynamical susceptibility $\Delta \chi^{\mathrm{so(dyn)}}(\mu)$ shows
a relatively large deviation from the pure Dirac case around the hybridization points $\epsilon_{1,2}$.
In particular, around $\epsilon_2$, the static susceptibility appears almost insensitive to the hybridization effect,
whereas the dynamical suceptibility
gets suppressed {by the hybridization}.
%  in comparison with the pure Dirac case.
This is because the dynamical susceptibility is dominated by the interband processes:
the contribution from the interband processes,
accompanied with the weight factor $F_{ab} = [f(\epsilon_a-\mu) - f(\epsilon_b-\mu)]/(\epsilon_a -\epsilon_b)$,
becomes significant at $\boldsymbol{k}$ around the hybridization point,
as the two bands $\epsilon_a$ and $\epsilon_b$ get close to one another.
As a result, the dynamical SO crossed susceptibility acquires a large modification from the band hybridization,
in comparison with the static susceptibility.

\begin{figure}[tbp]
    \includegraphics[width=8.4cm]{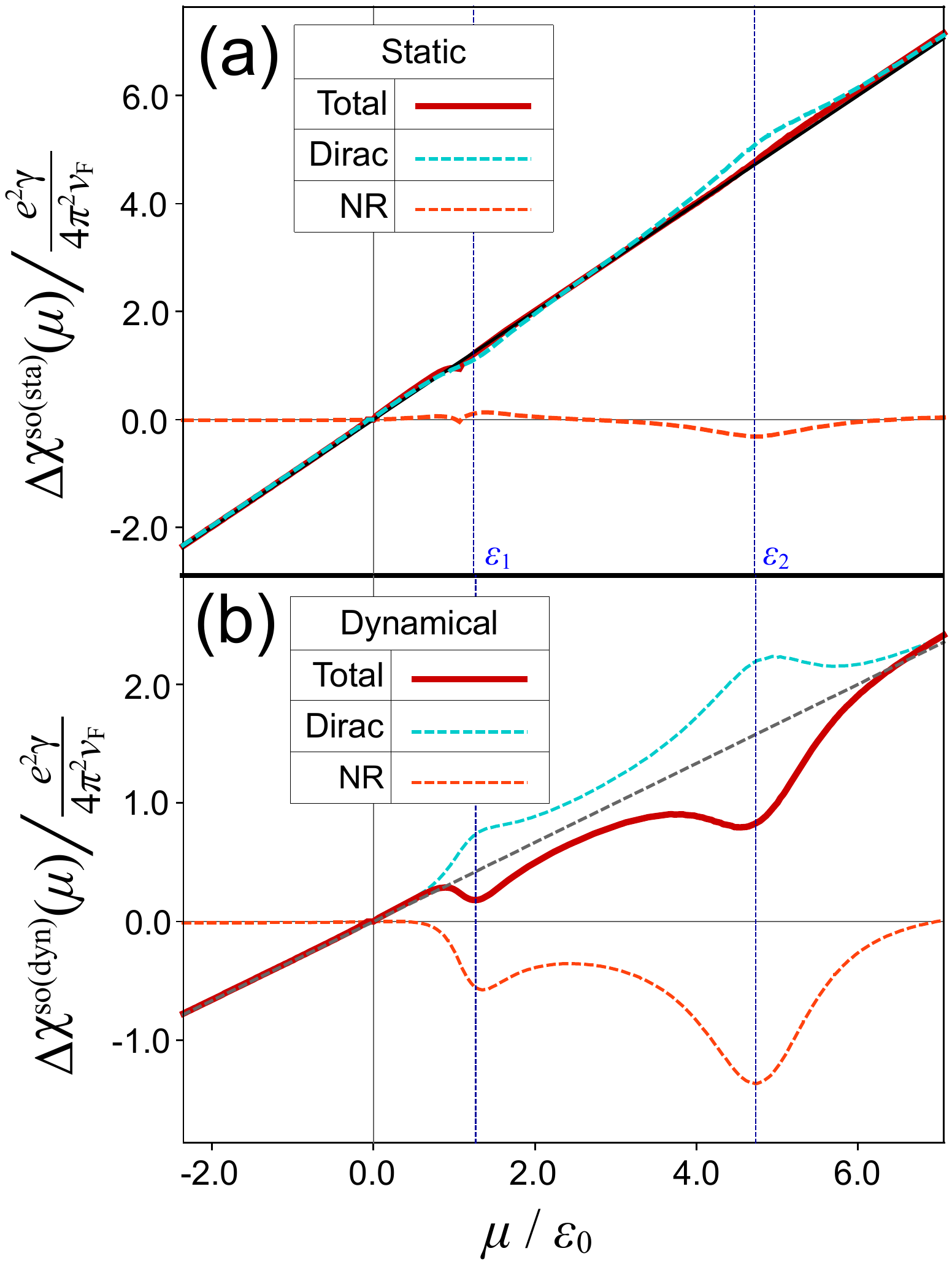}
    \caption{The SO crossed susceptibilities separated into the Dirac sector $\chi^{\mathrm{so}}_{\mathrm{Dirac}}$ and the nonrelativistic (NR) sector $\chi^{\mathrm{so}}_{\mathrm{NR}}$,
    and their sum (Total, $\chi^{\mathrm{so}}$).
    Panel (a) shows the static susceptibilities,
    while Panel (b) shows the dynamical susceptibilities.
    Both results are obtained with the offset of the NR band $\epsilon_0>0$, the effective mass for the NR band $m = 3\epsilon_0$, and the hybridization parameter $h=0.2\epsilon_0$.}
    \label{fig:hybb-pos-partial}
\end{figure}

\subsection{Response of Dirac and nonrelativistic sectors} \label{sec:channel-dependent}

In order to understand the hybridization-induced modification in $\chi^{\mathrm{so}}(\mu)$ in more detail,
we separate it into the {contributions from the} Dirac and nonrelativistic sectors.
The spin magnetization for the Dirac sector $\boldsymbol{M}^{\mathrm{s}}_{\mathrm{Dirac}}$
and that for the nonrelativistic sector $\boldsymbol{M}^{\mathrm{s}}_{\mathrm{NR}}$ can be
evaluated separately, with the spin operators
\begin{align}
    \bSpin_{\mathrm{Dirac}} = \frac{1}{2} 
    \begin{pmatrix}
        \bsig & 0 & 0 \\
        0 & \bsig & 0 \\
        0 & 0 & 0
    \end{pmatrix}, \quad
    \bSpin_{\mathrm{NR}} = \frac{1}{2} 
    \begin{pmatrix}
        0 & 0 & 0 \\
        0 & 0 & 0 \\
        0 & 0 & \bsig
    \end{pmatrix}.
\end{align}
As the response functions of these sector-resolved spin magnetizations
to the orbital magnetic field $\bB^{\mathrm{o}}$,
we define the SO crossed susceptibilities for the Dirac and nonrelativistic sectors,
\begin{align}
    M^{\mathrm{s}}_{\mathrm{Dirac},i} = -\gamma \langle S_{\mathrm{Dirac},i} \rangle &\equiv \chi^{\mathrm{so}}_{\mathrm{Dirac},ij} B^{\mathrm{o}}_j \\
    M^{\mathrm{s}}_{\mathrm{NR},i} = -\gamma \langle S_{\mathrm{NR},i} \rangle &\equiv \chi^{\mathrm{so}}_{\mathrm{NR},ij} B^{\mathrm{o}}_j,
\end{align}
which are obtained by using $\bSpin_{\mathrm{Dirac}}$ and $\bSpin_{\mathrm{NR}}$
instead of $\bSpin$ in the formulas shown in Sec.~\ref{sec:geometric-formula}.

Based on the above definition,
the sector-resolved susceptibilities as functions of the Fermi energy $\mu$ are obtained
as shown in Fig.~\ref{fig:hybb-pos-partial},
(a) in the static limit and (b) in the dynamical limit.
We here emphasize that $\chi^{\mathrm{so}}_{\mathrm{NR}}$ becomes nonzero around the hybridization points $\epsilon_{1,2}$
both in the static and dynamical limits,
among which the effect in the dynamical limit appears rather significant.
This indicates that the nonrelativistic sector shows a finite spin polarization in response to the orbital magnetic field,
even though the nonrelativistic fermions are originally not subject to SOC.

\modify{
    Let us here give a qualitative discussion about the mechanism
    how the dynamical susceptibility is strongly influenced by the band hybridization,
    by using a perturbation theory with a simple quantum mechanics.
    What we need to evaluate is the spin magnetic moment
    of the hybridized states around the hybridization points.
}
% Let us consider the behavior of the dynamical susceptibilities in more detail.
% Our calculation results indicate that the dynamical spin polarization in the Dirac sector
% gets enhanced by the hybridization,
% whereas the spin polarization in the nonrelativistic sector
% {is induced oppositely to the Dirac sector.}
% It can be qualitatively understood {by evaluating the spin magnetic moment}
% with the hybridized states around the hybridization points.
% Since Dirac fermions free from hybridization develops spin polarization in response to an orbital magnetic field,
% as seen in Section \ref{sec:weyl},
Since the orbital magnetic moment $\bmag_{aa}(\bk)$ of a massless Dirac (or Weyl) fermion
is parallel to its spin magnetic moment $\boldsymbol{\mu}_{aa}(\bk) = -\gamma \bSpin_{aa}(\bk)$,
as seen from Eqs.~(\ref{eq:Dirac-m}) and (\ref{eq:Dirac-spin}),
here we substitute the orbital magnetic field with
% we regard the effect of the orbital magnetic field as 
an effective spin magnetic field acting selectively on the Dirac sector,
coupled to the spins as $\delta \mathcal{H}_{\mathrm{eff}} = \gamma \boldsymbol{B}^{\mathrm{s}}_{\mathrm{Dirac}} \cdot \boldsymbol{S}_{\mathrm{Dirac}}$.
Note that this substitution of effective field is valid in describing only the direction of the induced spin polarization,
not its magnitude including parameter dependences.
We then take into account the hybridization between the Dirac and nonrelativistic sectors, and
consider the response of spin magnetic moments to this effective magnetic field.
For the sake of clarity,
we take the direction of $\boldsymbol{B}^{\mathrm{s}}_{\mathrm{Dirac}}$ as $z$-axis,
which yields $\delta \mathcal{H}_{\mathrm{eff}} = \gamma B^{\mathrm{s}}_{\mathrm{Dirac}} S^z_{\mathrm{Dirac}}$,
and focus on the magnetic moments in this direction,
as we are here interested in the responses in spatially isotropic systems.

At the momentum $\bk_c$ where the Dirac and nonrelativistic bands cross each other,
two states $|u_{\mathrm{Dirac}}(\bk_c)\rangle$ and $|u_{\mathrm{NR}}(\bk_c)\rangle$ get hybridized.
The hybridized states are given as linear combinations of the two states,
\begin{align}
    |u_\pm\rangle &= \tfrac{1}{\sqrt{2}} \left[ |u_{\mathrm{Dirac}}\rangle \pm |u_{\mathrm{NR}}\rangle\right], \label{eq:hyb-states}
\end{align}
{where the eigenenergies $\epsilon_\pm$ satisfy} $\epsilon_+ - \epsilon_- =2h$.
% (We here suppress the arguments $\bk_c$ for simplicity of notations.)
If the hybridization does not mix spins, as we have assumed {in the present model},
$|u_{\mathrm{Dirac}}\rangle$ and $|u_{\mathrm{NR}}\rangle$ participating to the hybridization
have the same spin polarizations.
If the Fermi level $\mu$ lies between $\epsilon_+$ and $\epsilon_-$,
only the occupied state $|u_-\rangle$ contributes to the spin polarization.
Therefore, we see here
how the spin magnetic moment of the state $|u_-\rangle$ gets perturbed by the effective magnetic field $\boldsymbol{B}^{\mathrm{s}}_{\mathrm{Dirac}}$.

When the effective magnetic field $\boldsymbol{B}^{\mathrm{s}}_{\mathrm{Dirac}}$ is applied,
the state $|u_-\rangle$ is perturbed by $\delta \mathcal{H}_{\mathrm{eff}}$,
\begin{align}
    |\delta u_-\rangle &= |u_+\rangle \frac{\langle u_+ | \delta \mathcal{H}_{\mathrm{eff}} |u_-\rangle }{\epsilon_- -\epsilon_+}
    = -\frac{\gamma B^{\mathrm{s}}_{\mathrm{Dirac}}}{2h}|u_+\rangle \langle u_+ | S^z_{\mathrm{Dirac}} |u_-\rangle
\end{align}
at first order in $\boldsymbol{B}^{\mathrm{s}}_{\mathrm{Dirac}}$.
By this perturbation, the spin magnetic moments of $|u_-\rangle$
projected onto the Dirac and nonrelativistic sectors,
which we denote as $\boldsymbol{\mu}_\Gamma = -\gamma \langle u_- | \boldsymbol{S}_\Gamma | u_-\rangle \ (\Gamma=\mathrm{Dirac},\mathrm{NR})$,
get modified as
\begin{align}
    \delta \mu^z_\Gamma &= -\gamma \delta \langle u_- | S^z_\Gamma | u_-\rangle 
    = -2\gamma  \mathrm{Re} \langle \delta u_- |S^z_\Gamma| u_-\rangle \nonumber \\
    &= \frac{\gamma^2 B^{\mathrm{s}}_{\mathrm{Dirac}}}{h} \mathrm{Re} \left[ \langle u_+ |S^z_\Gamma| u_-\rangle \langle u_- |S^z_{\mathrm{Dirac}}| u_+\rangle \right]. \label{eq:chi-XD}
\end{align}
For the Dirac sector, the modification
\begin{align}
    \delta \mu^z_{\mathrm{Dirac}} &= \frac{\gamma^2 B^{\mathrm{s}}_{\mathrm{Dirac}}}{h} |\langle u_+ |S^z_{\mathrm{Dirac}}| u_-\rangle|^2
\end{align}
is {parallel to $B^{\mathrm{s}}_{\mathrm{Dirac}}$},
from which we can qualitatively understand the enhancement of the SO response in the Dirac sector.
We note that
this mechanism is similar to the Van Vleck paramagnetism,
where the paramagnetic susceptibility is enhanced by the interband effect
that is allowed by SOC \cite{VanVleck}.
    On the other hand, for $\delta \mu^z_{\mathrm{NR}}$,
the matrix elements in Eq.~(\ref{eq:chi-XD}) become
\begin{align}
    \langle u_+ | S_{\mathrm{NR}}^z | u_- \rangle &= - \tfrac{1}{2} \langle u_{\mathrm{NR}} | S_{\mathrm{NR}}^z | u_{\mathrm{NR}} \rangle \\
    \langle u_- | S_{\mathrm{Dirac}}^z |u_+ \rangle &=  \tfrac{1}{2} \langle u_{\mathrm{Dirac}} | S_{\mathrm{Dirac}}^z | u_{\mathrm{Dirac}} \rangle,
\end{align}
    which yields
    \begin{align}
        \delta \mu^z_{\mathrm{NR}} &= -\frac{\gamma^2 B^{\mathrm{s}}_{\mathrm{Dirac}}}{4h} \langle u_{\mathrm{NR}} | S_{\mathrm{NR}}^z | u_{\mathrm{NR}} \rangle \langle u_{\mathrm{Dirac}} | S_{\mathrm{Dirac}}^z | u_{\mathrm{Dirac}} \rangle.
    \end{align}
Since we have assumed that 
$| u_{\mathrm{Dirac}} \rangle$ and $| u_{\mathrm{NR}} \rangle$ have the same spin direction,
$\langle u_{\mathrm{Dirac}} | S_{\mathrm{Dirac}}^z | u_{\mathrm{Dirac}} \rangle$
and $\langle u_{\mathrm{NR}} | S_{\mathrm{NR}}^i | u_{\mathrm{NR}} \rangle$ have the same signs,
which is the case with the present model Hamiltonian in Eq.~(\ref{eq:H-matrix}).
Therefore, the product of the two matrix elements in Eq.~(\ref{eq:chi-XD}) becomes negative,
yielding $\delta \mu^z_{\mathrm{NR}}$ {antiparallel to $B^{\mathrm{s}}_{\mathrm{Dirac}}$}.
This discussion provides qualitative interpretation about the negative SO response
induced in the nonrelativistic sector {seen in Fig.~\ref{fig:hybb-pos-partial}}(b),
which is due to {the structure of the hybridized states}.
% that the nonrelativistic sector develops spin polarization
% antiparallel to that in the Dirac sector.

\modify{
    Our calculation results of the SO crossed susceptibility using the minimal model in Eqs.~(\ref{eq:H-full}) and (\ref{eq:H-matrix})
    are well described by the above discussion with a simple quantum mechanics.
    This discussion is valid no matter what kind of internal degrees of freedom is present,
    such as orbital, sublattice, flavor, or color.
    Therefore, we can understand that the modifications in the SO susceptibilities found in our calculation
    are not the artifact from the minimal model employed here.
    It is the common feature available in any relativistic fermion systems
    hybridized with nonrelativistic fermion degrees of freedom,
    as long as the hybridization mixes two states with the same spin direction.
}

\section{Implication on experiments} \label{sec:experiments}
Finally, we give some discussions about
the implications of our findings on experiments.
In order to realize our idea in experimental measurements,
we first need to note the hierarchy of energy (time) scales,
among the relaxation rate (inverse relaxation time) $\tau^{-1}$,
the hybridization energy $h$,
and the frequency of the external magnetic field $\Omega$.
As mentioned in Sec.~\ref{sec:static-dynamical},
the static limit is valid for $\Omega < \tau^{-1} < h$,
and the dynamical limit {applies to the case} $\tau^{-1} < \Omega < h$.
For example,
the transport calculations in graphene with charged impurities
give the relaxation time around $\tau \sim 1\mathrm{ps}$,
corresponding to the frequency $\tau^{-1} \sim 1 \mathrm{THz}$ and the energy $4\mathrm{meV}$ \cite{Hwang_2008}.
This can be regarded as the typical scale of relaxation
for Dirac electrons at low carrier density,
with the Fermi velocity comparable to that of graphene ($\vf = 3\times 10^6\mathrm{m/s}$).
With this relaxation timescale,
transition between the static and dynamical behaviors in the susceptibility
can be achieved by varying the frequency of the external magnetic field
around the terahertz regime.
The hybridization energy $h$ should be {larger than} $\tau^{-1}$,
{so that the spectral broadening by the imaginary part of the fermion self energy may not} obscure
{the band splitting of $h$ arising from the hybridization effect}.

\subsection{Solid states}
For electrons in solid states,
the response to a magnetic field measured in experiments
contains both the response to the orbital magnetic field discussed throughout this article
and the response to the spin magnetic field via the Zeeman effect.
In order to extract the orbital effect,
one may excite the orbital degrees of freedom selectively
by a circularly polarized light,
and observe the magnetic circular dichroism,
namely the difference in the light absorption depending on the polarization of the light \cite{Thole_1992}.
Another way to identify the orbital effect
is to subtract the spin effect from the full response to the magnetic field.
In order to extract the spin effect,
one may rely on the exchange coupling between
the spins of localized electrons in magnetic elements and the spins of itinerant electrons,
which takes the same form with the Zeeman coupling.
By introducing magnetic dopants in bulk sample,
or the magnetic proximity effect in thin-film geometry attached with a magnetic material,
we can mimic the spin magnetic field for the itinerant electrons,
from which we may extract the response to the spin magnetic field.
% it is also important how to extract the SO crossed susceptibility in experiments.
% As we are interested in the response to the orbital magnetic field,
% we need to subtract {the response to} the spin magnetic field {via} the Zeeman effect
% {from the total susceptibility measured in response to the magnetic field.}
% While the spin and orbital magnetic fields cannot be tuned independently,
% the effect of spin magnetic field can be mimicked by exchange interaction,
% namely the coupling between the spins of an itinerant electron and a localized electron.
% It can be introduced by magnetic dopants in bulk sample,
% or the magnetic proximity effect in thin-film geometry attached with a magnetic material.
% Another possible solution is to excite the orbital degrees of freedom selectively
% by a circularly polarized light,
%     and observe the magnetic circular dichroism,
%     namely the difference in the light absorption depending on the polarization of the light \cite{Thole_1992}.
% or by mechanical rotation to observe the Barnett effect.

Aside from the total spin polarization,
we are also interested in the spin polarization separated into the Dirac and nonrelativistic sectors,
as discussed in Sec.~\ref{sec:channel-dependent}.
In order to distinguish the spin polarization by the sectors,
the nuclear magnetic resonance (NMR) spectroscopy will be helpful in some materials \cite{Mehring}.
{
    We may rely on the Knight shift in the NMR spectrum,
    which arises from the hyperfine coupling bewteen the electron spin and the nuclear magnetic moment \cite{Knight_1949}.
    The Knight shift provides information about
    the electron spin polarization belonging to each constituent element in the compound.
    Therefore, if the Dirac and nonrelativistic bands in the material come from different elements,
    such as in a Dirac semimetal with impurity dopants, 
    the Knight shift may provide information about the sector-resolved spin polarization mentioned in our discussion.
}

\subsection{Quark matter}
Our discussion can also be applied to quark matter.
In particular, we may consider mixture of heavy quarks, corresponding to the flavor $c$ and sometimes $b$,
and light quarks $u$, $d$, and $s$.
Light quarks,
having Dirac masses relatively smaller than heavy quarks,
can be treated as Dirac fermions,
while heavy quarks behave as nonrelativistic fermions at low momentum.
Throughout a relativistic heavy-ion collision process,
the generated quark matter
will be subject to a magnetic field,
if the collision of two heavy nuclei is noncentral \cite{Kharzeev_2008,Rafelski_1976,Kharzeev_2016,Huang_2016,Zhao_2019}.
In a {manner similar to} the CME and the CSE proposed in quark matter,
this magnetic field will give rise to the spin polarization of {both the light and heavy} quarks.
% As quarks are described as relativistic Dirac fermions,
% the magnetic field couples to the quarks only via the vector potential,
% not by the Zeeman coupling.
Since light quarks are described as relativistic Dirac fermions,
the magnetic field couples to them only via the vector potential.
While heavy quarks behave as nonrelativistic fermions,
the Zeeman effect on them is almost negligible due to their large Dirac masses.
Therefore, {the effect of the magnetic field on the spin polarization of quarks} can be
{dominantly} described by the SO crossed susceptibility $\chi^{\mathrm{so}}$.

Hybridization of light and heavy quarks is possible,
if heavy quarks are dilute enough in comparison with light quarks.
Heavy quarks form bound states with light quarks by the strong interaction,
which is proposed as the QCD Kondo effect \cite{Yasui_2013,Yasui_2016,Yasui_2017c,Yasui_2019c,Hattori_2015,Ozaki_2016,Yasui_2019,Yasui_2017,Kanazawa_2016,Kimura_2017,Yasui_2017a,Suzuki_2017,Yasui_2017b,Kimura_2019,Fariello_2019,Hattori_2019,Suenaga_2020,Suenaga_2020a,Kanazawa_2020,Araki_2021}.
Under such a hybridization,
our analysis on the SO susceptibility implies that heavy quarks,
corresponding to the nonrelativistic sector in our analysis,
develop spin polarization in response to a magnetic field,
although the Zeeman coupling for heavy quarks is weak.
While the spin polarization of heavy quarks cannot be measured directly,
it may be captured as spin polarization of heavy hadrons including heavy quarks ($c$ or $b$) after hadronization,
where the quarks are cooled down and confined in hadrons.
In the hadronization process,
spin polarization of heavy quark can be transferred
to spin polarization of a $\Lambda_c$ or $\Lambda_b$ baryon.
% in the heavy-quark mass limit.
Therefore,
measurement of the spin polarization of the $\Lambda_c$ or $\Lambda_b$ baryon
is one of the promising ways
to observe the hybridization induced by the QCD Kondo effect
which is not experimentally verified so far.
In order to understand such an effect in quark matter precisely,
one needs to determine the microscopic structure of interaction and parameters specific to the system,
which is left for further analysis \cite{Suenaga_CSE}.

\section{Conclusion} \label{sec:conclusion}
In the present article,
we have focused on the SO crossed susceptibility,
namely the response function of the spin magnetization $\boldsymbol{M}^{\mathrm{s}}$ {composed of} the spin polarization {of fermions},
to the orbital magnetic field $\bB^{\mathrm{o}}$ described by the U(1) vector potential.
The SO crossed susceptibility quantifies the relativistic effect acting on fermions,
since it arises as a consequence of SOC,
which is the relativistic effect.
The idea of SO susceptibility is applicable to any {kind} of fermion system,
not limited to solid states but also to quark matter.

One of the main issues discussed in this work is the comparison of the SO crossed susceptibilities in two limits,
namely the static and dynamical limits.
% These two limits are distinguished by the time scale in introducing the magnetic field,
% which matters how the fermion distribution gets modified by the magnetic field.
While the behavior of the SO crossed susceptibility in the static limit,
induced by a slowly introduced magnetic field keeping the equilibrium,
is broadly discussed in the context of topological materials,
its dynamical-limit behavior,
under an abruptly introduced magnetic field that drives the distribution out of equilibrium,
is discussed systematically for the first time.

% The general forms for the crossed susceptibilities in the static and dynamical limits
% are obtained as shown in Section \ref{sec:geometric-formula},
% using the geometric quantities in momentum space,
% namely the Berry connection $\bA$, the Berry curvature $\boldsymbol{\Omega}$,
% the spin Berry curvature $\boldsymbol{\Omega}^{(\Spin_i)}$,
% and the orbital magnetic moment of a single particle $\boldsymbol{m}^{\mathrm{orb}}$.

As a result of our analysis,
we have found that the difference between the static and dynamical SO susceptibilities becomes significant
in the presence of band hybridization.
We have seen this tendency by using the hybridized model of
Dirac fermions obeying spin-momentum locking and nonrelativistic fermions free from SOC.
% Using the obtained formulae,
% the SO crossed susceptibility is evaluated for the hybridized system composed of
% the Dirac fermions obeying spin-momentum locking and the nonrelativistic fermions free from spin-orbit coupling.
% As a result,
% we have demonstrated that the difference between the static and dynamical susceptibilities becomes qualitatively clear in the vicinity of the band hybridization point.
In the dynamical limit,
the SO susceptibility gets strongly modified by the hybridization,
and the spins of the nonrelativistic fermions also respond to the orbital magnetic field,
even though they are not originally subject to SOC.
These modification effects can be understood as the outcome of interband perturbation effect allowed by the band hybridization,
which is in a {mechanism similar to} the Van Vleck paramagnetism.
% While the Dirac fermions exhibit a universal behavior in the crossed susceptibility
% due to their spin-momentum locking feature,
% we have seen here that the spins of the nonrelativistic fermions also respond to the orbital magnetic field in the dynamical limit,
% driven by the interband processes.
% In other words, one can say that the relativistic effect (spin-orbit coupling) on the Dirac fermions is partially transferred to the nonrelativistic fermions,
% due to the hybridization between them.

The framework of our discussion applies at various energy scales,
such as electrons in solids and quark matter after heavy-ion collision in accelerators.
% Experimental realization of the crossed responses in such systems is thoroughly discussed in {Section \ref{sec:experiments}}.
\modify{
    In the present article, we have taken a simple model with minimal number of degrees of freedom,
    with which we are successful in extracting a common feature in the SO susceptibility
    triggered by the hybridization of relativistic and nonrelativistic degrees of freedom.
    Of course, in realistic systems,
    there may be more diverse internal degrees of freedom,
    depending on the atomic orbitals and crystalline symmetries for electrons in solid states
    and color and flavor degrees of freedom for quarks in high-energy physics.
}
Detailed treatment of the crossed response phenomena,
based on the microscopic model for each setup,
will be left for future analysis.

\acknowledgments{
    Y.~A. is supported by the Leading Initiative for Excellent Young Researchers (LEADER).
    D.~S. wishes to thank Keio University and Japan Atomic Energy Agency (JAEA) for their hospitalities during his stay there.
    K.~S. is supported by Japan Society for the Promotion of Science (JSPS) KAKENHI
    (Grants No.~JP17K14277 and JP20K14476).
    S.~Y. is supported by JSPS KAKENHI (Grant No.~JP17K05435)
    and by the Interdisciplinary Theoretical and Mathematical Sciences Program (iTHEMS) at RIKEN.
    }

\appendix

\section{Orbital and spin magnetizations} \label{sec:orbital-spin} 
In this part of the appendix,
we review how a magnetic field couples to the orbital and spin degrees of freedom,
and distinguish the SO crossed susceptibility from the other types of magnetic susceptibilities.

For relativistic fermions under Lorentz symmetry,
their coupling to a magnetic field $\bB$ is given in terms of covariant derivative,
with the vector potential $\bA(\br)$ corresponding to the magnetic field $\bB = \bnab \times \bA$.
On the other hand, for nonrelativistic fermions,
such as electrons trapped in crystals
(including electrons in topological semimetals with the Dirac- or Weyl-type dispersion at low energy),
their coupling to a magnetic field is classified into (a) the orbital effect and (b) the spin effect,
which can be derived by downfolding the relativistic theory to the nonrelativistic limit.

(a) The \textit{orbital effect} is given in terms of covariant derivative,
which is similar to the gauge coupling in the relativistic theory.
If the dynamics of fermions is described by the continuum Hamiltonian
\begin{align}
    \mathcal{H}_0 &= \int d\br \ \psi^\dag(\br) H(\bp) \psi(\br), 
\end{align}
where $\psi(\br)$ is the (multi-component) field operator of the fermions
and $H(\bp)$ is defined by substituting the momentum operator $\bp = -i\boldsymbol{\nabla}$ to the momentum-space Hamiltonian $H(\bk)$,
the {vector potential} shifts the Hamiltonian as
\begin{align}
    \mathcal{H} = \int d\br \ \psi^\dag(\br) H(\bp - e\bA(\br)) \psi(\br),
\end{align}
with $-e$ the electric charge of the fermion.
Therefore, the perturbation by the magnetic field is extracted as
\begin{align}
    \delta \mathcal{H}^{\mathrm{o}} &= -\frac{e}{2} \int d\br \ \psi^\dag(\br) \{ \bv(\bp), \bA(\br) \} \psi(\br), \label{eq:mag-orbital}
\end{align}
with the velocity operator $\bv(\bp) = \partial H(\bp)/\partial \bp$.

(b) The \textit{spin effect} is the so-called Zeeman splitting,
namely the coupling between the spin angular momentum and the magnetic field.
If the spin density operator of the fermions is given as $\boldsymbol{S} = \psi^\dag \bSpin \psi$,
with $\bSpin$ the matrix acting on the components of the field operator,
the spin effect of the magnetic field is given in terms of the Zeeman term,
\begin{align}
    \delta \mathcal{H}^{\mathrm{s}} &= \gamma \int d\br \ \bB \cdot \boldsymbol{S} = \gamma \int d\br \ \psi^\dag (\bB \cdot \bSpin) \psi.
\end{align}
Here $\gamma = g\mu_B$ is the parameter called gyromagnetic ratio,
with $\mu_B$ the Bohr magneton and $g$ is the $g$-factor for the fermions.
Note that $g$ may depend on internal degrees of freedom,
such as the species of fermions, the atomic orbitals that the electrons in the crystal belong to, etc.,
whereas we here neglect such detailed structures.

Although the origins of the above two effects are the same magnetic field $\bB$,
one may formally distinguish them by using different labels for the magnetic field, $\bB^{\mathrm{o}}$ and $\bB^{\mathrm{s}}$.
Under the orbital and spin magnetic fields,
the partition function $Z[\bB^{\mathrm{o}},\bB^{\mathrm{s}}]$ is given
from the perturbed Hamiltonian $\mathcal{H}_0 + \delta\mathcal{H}^{\mathrm{o}}[\bB^{\mathrm{o}}] + \delta\mathcal{H}^{\mathrm{s}}[\bB^{\mathrm{s}}]$
by tracing out the fermionic degrees of freedom $(\psi^\dag,\psi)$.
From this partition function,
the magnetization can also be defined separately for the orbital and spin sectors,
as {thermodynamic variables} conjugate to $\bB^{\mathrm{o}}$ and $\bB^{\mathrm{s}}$:
\begin{align}
    \boldsymbol{M}^{\mathrm{o}} = -\frac{\delta\ln Z}{\delta \bB^{\mathrm{o}}} \biggr|_{\bB^{\mathrm{o}}=\bB^{\mathrm{s}}=0}, \ 
    \boldsymbol{M}^{\mathrm{s}} = -\frac{\delta\ln Z}{\delta \bB^{\mathrm{s}}} \biggr|_{\bB^{\mathrm{o}}=\bB^{\mathrm{s}}=0}.
    \nonumber
\end{align}
The spin magnetization $\boldsymbol{M}^{\mathrm{s}}$ is composed of the spins of the fermions,
related to the expectation value of spin polarization $\langle \boldsymbol{S} \rangle$ as
\begin{align}
    \boldsymbol{M}^{\mathrm{s}} = -\gamma \langle \boldsymbol{S} \rangle = -\gamma \langle \psi^\dag \bSpin \psi \rangle.
\end{align}
On the other hand, the orbital magnetization $\boldsymbol{M}^{\mathrm{o}}$ comes from the orbital angular momenta of the fermions,
corresponding to the circulating electric current carried by the fermions.
Since the position operator $\br$ is {ill-defined} in unbounded systems,
the {momentum-space formalism of} orbital magnetization $\boldsymbol{M}^{\mathrm{o}}$ cannot be given
{so simply} as the spin magnetization $\boldsymbol{M}^{\mathrm{s}}$
[see Eq.~(\ref{eq:M-orbital})].

The magnetic susceptibility $\chi_{ij}$ is defined as the tensor characterizing
the response of magnetization $\delta M_i$ to the magnetic field $B_j$.
As the magnetic field and the magnetization are separated into the orbital and spin parts defined above,
the magnetic susceptibility can be separated into the four parts:
\begin{align}
    \text{[Spin-spin]} \quad & \delta M^{\mathrm{s}}_i = {\chi}^{\mathrm{ss}}_{ij} B^{\mathrm{s}}_j, \\
    \text{[Spin-orbital]} \quad & \delta M^{\mathrm{s}}_i = {\chi}^{\mathrm{so}}_{ij} B^{\mathrm{o}}_j, \\
    \text{[Orbital-spin]} \quad & \delta M^{\mathrm{o}}_i = {\chi}^{\mathrm{os}}_{ij} B^{\mathrm{s}}_j, \\
    \text{[Orbital-orbital]} \quad & \delta M^{\mathrm{o}}_i = {\chi}^{\mathrm{oo}}_{ij} B^{\mathrm{o}}_j. 
\end{align}
The spin-spin response is known as the Pauli paramagnetism,
namely, the spin polarization induced by the Zeeman splitting,
and the orbital-orbital part often gives rise to the Landau diamagnetism,
due to the orbital magnetic moment of the quantum Hall states under the Landau quantization.
The spin-orbital crossed parts $\chi^{\mathrm{so}}$ and $\chi^{\mathrm{os}}$ are not classified with either of them,
which require the correlation {between} the spin and orbital degrees of freedom.

The above four susceptibilities are given in terms of the partition function $Z[\bB^{\mathrm{o}},\bB^{\mathrm{s}}]$,
\begin{align}
    \chi^{\mathrm{ss}}_{ij} = -\frac{\delta^2 \ln Z}{\delta B^{\mathrm{s}}_i \delta B^{\mathrm{s}}_j}, &\quad
    \chi^{\mathrm{so}}_{ij} = -\frac{\delta^2 \ln Z}{\delta B^{\mathrm{s}}_i \delta B^{\mathrm{o}}_j}, \\
    \chi^{\mathrm{os}}_{ij} = -\frac{\delta^2 \ln Z}{\delta B^{\mathrm{o}}_i \delta B^{\mathrm{s}}_j}, &\quad
    \chi^{\mathrm{oo}}_{ij} = -\frac{\delta^2 \ln Z}{\delta B^{\mathrm{o}}_i \delta B^{\mathrm{o}}_j}.
\end{align}
The spin-orbital crossed parts $\chi^{\mathrm{so}}$ and $\chi^{\mathrm{os}}$ satisfy the relation
\begin{align}
    \chi^{\mathrm{so}}_{ij} = \chi^{\mathrm{os}}_{ji}, \label{eq:Onsager}
\end{align}
which is the outcome of the Onsager's reciprocity theorem \cite{Onsager}.

    Although magnetic field and magnetization in the relativistic regime cannot be separated into the orbital and spin parts,
    the idea of the SO crossed susceptibility $\chi^{\mathrm{so}}$ still applies.
    The spin magnetization can be defined from the spin polarization,
    $\boldsymbol{M}^{\mathrm{s}} = -\gamma \langle \boldsymbol{S} \rangle$,
    and the magnetic field $\boldsymbol{B}$ couples to the particles only via the vector potential.
    Therefore, in the relativistic regime,
    the response of the spin magnetization to the magnetic field
    is described in terms of $\chi^{\mathrm{so}}$ defined above.

\section{Detailed derivation process of the SO susceptibility tensor} \label{sec:derivation}
In this appendix,
we show details of the derivation process toward the formula for $\chi^{\mathrm{so}}$,
whose final form is given in Sec.~\ref{sec:geometric-formula}.

\subsection{Perturbation by vector potential}
The starting point is the perturbation by {the coupling to} the vector potential $\bA$.
With the perturbation Hamiltonian $\delta H^{\mathrm{o}}$ given by Eq.~(\ref{eq:perturbation-orbital}),
the linear perturbation of the Green's function becomes
\begin{align}
    & \delta G (\bk,i\omega_n ; \bk',i\omega_n') \\
    & = G(\bk,i\omega_n) \delta H^{\mathrm{o}}(\bk,i\omega_n;\bk',i\omega_n') G(\bk',i\omega_n'), \nonumber
\end{align}
which is nondiagonal in momentum and Matsubara frequency.
Using this perturbation of Green's function,
the expectation value of the spin polarization $\langle \boldsymbol{S} (\bq,i\bar{\omega}_m) \rangle$
induced by the vector potential $\bA(\bq,i\bar{\omega}_m)$ reads
\begin{align}
    & \langle S_i(\bq,i\bar{\omega}_m) \rangle \nonumber \\
    &= \int d\br d\tau \ e^{i\bq\cdot\br + i\bar{\omega}_m \tau} \langle S_i(\br,\tau) \rangle\\
    &= \int d\br d\tau \ e^{i\bq\cdot\br + i\bar{\omega}_m \tau} \langle \psi^\dag(\br,\tau) \Spin_i \psi(\br,\tau) \rangle \\
    &= \int \frac{d\br d\tau}{(\beta V)^2} \sum_{i\omega_n,i\omega'_{n}} \sum_{\bk,\bk'} e^{i(\bq+\bk'-\bk)\cdot\br + i(\bar{\omega}_m+\omega'_n-\omega_n)\tau} \nonumber \\
    & \hspace{3.5cm} \times \langle \psi^\dag(\bk',i\omega'_n) \Spin_i \psi(\bk,i\omega_n) \rangle
\end{align}
\begin{align}
    &= \frac{1}{\beta V} \sum_{i\omega_n,\bk} \left\langle \psi^\dag(\bk-\bq,i\omega_n-i\bar{\omega}_m) \Spin_i \psi(\bk,i\omega_n) \right\rangle \\
    &= \frac{-1}{\beta V} \sum_{i\omega_n,\bk} \Tr \left[ \Spin_i \delta G(\bk,i\omega_n ; \bk-\bq,i\omega_n-i\bar{\omega}_m) \right]  \\
    &= \frac{-1}{\beta V} \sum_{i\omega_n,\bk} \Tr \bigl[ \Spin_i G(\bk,i\omega_n) \delta H (\bk,\bk-\bq) \\
    & \quad \quad \quad \quad \quad \quad \quad \quad \times G(\bk-\bq,i\omega_n-i\bar{\omega}_m) \bigr] \nonumber
\end{align}
\begin{align}
    &= -\frac{e A_l(\bq,i\bar{\omega}_m)}{2\beta V} \sum_{i\omega_n,\bk} \Tr \bigl\{ \Spin_i G(\bk,i\omega_n) \\
    & \quad \quad \quad \quad \times \left[v_l (\bk)+v_l(\bk-\bq)\right] G(\bk-\bq,i\omega_n-i\bar{\omega}_m) \bigr\} \nonumber \\
    & \equiv \Pi_{A_l}^{\Spin_i}(\bq,i\bar{\omega}_m) A_l(\bq,i\bar{\omega}_m),
\end{align}
where $i\bar{\omega}_m = i\tfrac{2\pi}{\beta}m$ is the bosonic Matsubara frequency
corresponding to the frequency of the vector potential.
This form corresponds to Eq.~(\ref{eq:perturbed-spin}).
By the {analytical continuation} $i\bar{\omega}_m \rightarrow \Omega+i0$,
we obtain the linear response to the vector potential,
\begin{align}
    \langle S_i(\bq,\Omega) \rangle &= \Pi_{A_l}^{S_i}(\bq,\Omega) A_l(\bq,\Omega),
\end{align}
for finite momentum $\bq$ and frequency $\Omega$.

\subsection{Description with band eigenstates}
We need to evaluate the response tensor $\Pi_{A_l}^{S_i}(\bq,\Omega)$ up to $O(\bq)$,
to apply Eq.~(\ref{eq:response-chi-so}).
By decomposing the Green's function as
\begin{align}
    G(\bk,i\omega_n) &= -\sum_{a} \frac{|u_{a}(\bk)\rangle \langle u_{a}(\bk)|}{i\omega_n^+ - \epsilon_a(\bk)},
\end{align}
where $a$ denotes the band index satisfying
\begin{align}
    H(\bk) |u_a(\bk)\rangle = \epsilon_a(\bk) |u_a(\bk)\rangle
\end{align}
and $i\omega_n^+ = i\omega_n +\mu$,
the response tensor is given in terms of matrix elements as
\begin{align}
    & \Pi_{A_l}^{S_i}(\bq,i\bar{\omega}_m) \nonumber \\
    &= \frac{-e}{2 \beta V} \sum_{i\omega_n,\bk} \Tr \Bigl\{ \Spin_i G(\bk,i\omega_n) \\
    & \hspace{1cm} \times \left[v_l (\bk)+v_l(\bk-\bq)\right] G(\bk-\bq,i\omega_n-i\bar{\omega}_m) \Bigr\} \nonumber \\
    &= \frac{-e}{2 \beta V} \sum_{i\omega_n,\bk} \sum_{ab} \frac{ \langle u_b(\bk-\bq) | \Spin_i |  u_a(\bk) \rangle }{\left[i\omega_n^+ - \epsilon_a(\bk)\right] \left[i\omega_n^+ -i\bar{\omega}_m - \epsilon_b(\bk-\bq)\right]} \nonumber \\
    & \hspace{1cm} \times \langle u_a(\bk) | v_l (\bk)+v_l(\bk-\bq) | u_b(\bk-\bq)\rangle \\
    & \equiv \frac{-e}{\beta V} \sum_{i\omega_n,\bk} \sum_{ab} \frac{ \MM_{ab}^{il}(\bk,\bq) }{\left[i\omega_n^+ - \epsilon_a(\bk)\right] \left[i\omega_n^+ -i\bar{\omega}_m - \epsilon_b(\bk-\bq)\right]}.
\end{align}
Here $\MM_{ab}^{il}(\bk,\bq)$ defined in the last line {corresponds to} Eq.~(\ref{eq:response-M}).

The Matsubara summation is evaluated as
\begin{align}
    & \frac{1}{\beta} \sum_{i\omega_n} \frac{1}{\left[i\omega_n^+ - \epsilon_a(\bk) \right] \left[i\omega_n^+ -i\bar{\omega}_m - \epsilon_b(\bk-\bq)\right]} \nonumber\\
    &= \frac{f(\epsilon_a(\bk)-\mu) - f(\epsilon_b(\bk-\bq)-\mu +i\bar{\omega}_m)}{-i\bar{\omega}_m +\epsilon_a(\bk) -\epsilon_b(\bk-\bq)} \\
    &= \frac{f(\epsilon_a(\bk)-\mu) - f(\epsilon_b(\bk-\bq)-\mu)}{-i\bar{\omega}_m +\epsilon_a(\bk) -\epsilon_b(\bk-\bq)} \\
    &\overset{i\bar{\omega}_m \rightarrow \Omega+i0}{\longrightarrow} \frac{f(\epsilon_a(\bk)-\mu) - f(\epsilon_b(\bk-\bq)-\mu)}{\epsilon_a(\bk) -\epsilon_b(\bk-\bq) -\Omega-i0} \equiv \FF_{ab}(\bk,\bq,\Omega), \nonumber
\end{align}
which corresponds to Eq.~(\ref{eq:response-F}).
We thus obtain Eq.~(\ref{eq:response-Pi}),
\begin{align}
    \Pi^{S_i}_{A_l}(\bq,\Omega) &= -\frac{e}{V}\sum_{\bk} \sum_{ab} \FF_{ab}(\bk,\bq,\Omega) \MM_{ab}^{il}(\bk,\bq).
\end{align}

\subsection{Expansion by $\bq$}
We need to expand $\Pi_{A_j}^{\mathcal{O}}(\bq,\Omega)$ up to $O(\bq)$
to derive the response to the magnetic field.
By expanding $\FF_{ab}(\bk,\bq,\Omega)$ and $\MM_{ab}^{il}(\bk,\bq)$ as
\begin{align}
    \FF_{ab}(\bk,\bq,\Omega) &= \FF_{ab}^{(0)}(\bk,\Omega) - q_h \FF_{ab}^{(1) h}(\bk,\Omega) + O(q^2), \\
    \MM_{ab}^{il}(\bk,\bq) &= \MM_{ab}^{(0)il}(\bk) - q_h \MM_{ab}^{(1)ilh}(\bk) + O(q^2),
\end{align}
$\chi^{\mathrm{so}}_{ij}(\bq = 0,\Omega)$ can be obtained from Eq.~(\ref{eq:response-chi-so}) as
\begin{align}
    & \chi^{\mathrm{so}}_{ij}(\bq =0,\Omega) = \frac{i}{2}\gamma \epsilon_{jlh} \frac{\partial \Pi^{S_i}_{A_l}(\bq,\Omega)}{\partial q_h} \biggr|_{\bq =0} \label{eq:expansion} \\
    & \hspace{1cm} = -\frac{ie}{2V} \epsilon_{jlh} \sum_{\bk} \sum_{ab} \left[ \FF_{ab}^{(1)h} \MM_{ab}^{(0)il} + \FF_{ab}^{(0)} \MM_{ab}^{(1)ilh} \right]. \nonumber
\end{align}

The expansion of the matrix elements $\MM_{ab}^{il}(\bk,\bk-\bq)$ is given by
\begin{align}
    \MM_{ab}^{(0)il} &= \langle u_a | v_l | u_b \rangle \langle u_b | \Spin_i | u_a \rangle, \\
    \MM_{ab}^{(1)ilh} &= \tfrac{1}{2} \langle u_a | \partial_{k_h} v_l | u_b \rangle \langle u_b | \Spin_i | u_a \rangle \\
    & \quad + \langle u_a | v_l | \partial_{k_h} u_b\rangle \langle u_b | \Spin_i | u_a \rangle \nonumber \\
    & \quad + \langle u_a | v_l | u_b \rangle \langle \partial_{k_h} u_b | \Spin_i | u_a \rangle, \nonumber
\end{align}
where the dependence on $\bk$ is not explicitly written for simplicity.
$v_a^h$ is the group velocity, defined by $v_a^h = \partial_{k_h} \epsilon_a$.
Since the first term in $\MM_{ab}^{(1)ilh}$ contains $\partial_{k_h} v_l = \partial_{k_h} \partial_{k_l} H(\bk)$,
which is symmetric in $h \leftrightarrow l$,
it does not contribute to $\chi^{\mathrm{so}}_{ij}(\bq,\Omega)$ after the antisymmetrization by $\epsilon_{jlh}$.

On evaluating the weight factor $\FF_{ab}(\bk,\bq,\Omega)$,
we should be careful about the difference between the static and dynamical limits,
which comes from the absence or presence of the frequency $\Omega$ in the denominator.
\begin{itemize}
    \item For the \textit{interband} process $\epsilon_a \neq \epsilon_b$,
    the numerator and the denominator remain finite in the limit $\bq \rightarrow 0$ and $\Omega \rightarrow 0$
    irrespective of the order of taking these two limits,
    and hence the $\bq$-expansion of $\FF_{ab}(\bk,\bq,\Omega \rightarrow 0)$ is straightforwardly given as
    \begin{align}
        \FF_{ab}^{(0)} &= \frac{f_a - f_b}{\epsilon_a - \epsilon_b}, \label{eq:F-expansion} \\
        \FF_{ab}^{(1) h} &= v_a^h \left[ \frac{f_a - f_b}{(\epsilon_a - \epsilon_b)^2} - \frac{f'_a}{\epsilon_a - \epsilon_b} \right],
        \nonumber
    \end{align}
    in both the static and dynamical limits.
    \item For the \textit{intraband} process $\epsilon_a = \epsilon_b$,
    the denominator approaches zero in the limit $\bq \rightarrow 0$ and $\Omega \rightarrow 0$,
    and hence we need to care about the order of taking these two limits.
    In the static limit, where the limit $\Omega = 0$ is taken first,
    both the numerator and the denominator approach zero in the limit $\bq \rightarrow 0$,
    yielding
    \begin{align}
        & \FF_{ab}(\bk,\bq,\Omega=0)|_{\epsilon_a = \epsilon_b} \\
        &= \frac{f(\epsilon_a(\bk)) - f(\epsilon_a(\bk-\bq))}{\epsilon_a(\bk) - \epsilon_a(\bk-\bq)} \overset{\bq\rightarrow 0}{\longrightarrow} f'(\epsilon_a(\bk)) = f'_a,
        \nonumber
    \end{align}
    which is of $O(q^0)$.
    In the dynamical limit, where $\Omega$ is kept finite first,
    the denominator remains finite in the limit $\bq \rightarrow 0$.
    Therefore, the $\bq$-expansion becomes
    \begin{align}
        & \FF_{ab}(\bk,\bq,\Omega)|_{\epsilon_a = \epsilon_b} \\
        &= \frac{f(\epsilon_a(\bk)) - f(\epsilon_a(\bk-\bq))}{\epsilon_a(\bk) - \epsilon_a(\bk-\bq) -\Omega} 
        = -\frac{q_h v_a^h f'_a}{\Omega} + O(q^2), \nonumber
    \end{align}
    which is of $O(q^1)$.
    Although it appears to diverge in the limit $\Omega \rightarrow 0$,
    it does not contribute to $\chi^{\mathrm{so(dyn)}}$ after the antisymmetrization,
    as we shall see below.
\end{itemize}

\subsection{Rearrangement with geometric quantities} \label{sec:geometric}
From the $\bq$-expansion given by Eq.~(\ref{eq:expansion}),
we are now ready to rearrange the obtained terms into the geometric quantities,
\begin{align}
    \bA_{ab} &= i \langle u_a | \bnab_k u_b \rangle, \\
    \bmag_{ab} &= \frac{ie}{2} \langle \bnab_k u_a | \times ( \bar{\epsilon}_{ab} - H ) | \bnab_k u_{b} \rangle, \\
    \bOm_{ab} &= i \langle \bnab_k u_a | \times | \bnab_k u_b \rangle, \\
    \bOm^{(\Spin_i)}_{ab} &= i \langle \bnab_k u_a | \times \Spin_i | \bnab_k u_b \rangle,
\end{align}
whose physical meanings are briefly explained in Sec.~\ref{sec:geometric-formula}.
We also use the shorthand notations
\begin{align}
    \bar{\epsilon}_{ab} &= \tfrac{1}{2} (\epsilon_a + \epsilon_b), \\
    \Spin^i_{ba} &= \langle u_b | \Spin_i | u_a \rangle.
\end{align}
The matrix element of the velocity operator can be transformed as
\begin{align}
    & \langle u_a | \boldsymbol{v} |u_b \rangle = \langle u_a | \bnab_k H | u_b \rangle \\
    &= \bnab_k \langle u_a | H | u_b \rangle - \langle \bnab_k u_a | H | u_b \rangle - \langle u_a | H | \bnab_k u_b \rangle \nonumber \\
    &= \bnab_k \epsilon_a \delta_{ab} - \epsilon_b \langle \bnab_k u_a  | u_b \rangle - \epsilon_a \langle u_a | \bnab_k u_b \rangle \nonumber \\
    &= \bv_a \delta_{ab} +(\epsilon_b - \epsilon_a) \langle u_a | \bnab_k u_b \rangle. \nonumber
\end{align}
Therefore, we obtain the relation
\begin{align}
    \bv | u_a \rangle &= \bv_a |u_b\rangle + (\epsilon_a -H) |\bnab_k u_a\rangle .
\end{align}

\subsubsection{Intraband contribution}
The intraband contribution differs in the static and dynamical limits.
In the static limit, the weight factor is
\begin{align}
    \FF^{(0)}_{ab}|_{\epsilon_a = \epsilon_b} &= f'_a,
\end{align}
and hence the intraband contribution to the susceptibility becomes
\begin{align}
    \chi^{\mathrm{so(sta:intra)}}_{ij} &= -\frac{ie}{2V} \epsilon_{jlh} \sum_{\bk} \sum_{a \equiv b} f'_a \MM_{ab}^{(1)ilh}.
\end{align}
Here $\MM_{ab}^{(1)ilh}$ for $\epsilon_a = \epsilon_b$ reads (by omitting the term symmetric in $l \leftrightarrow h$)
\begin{align}
    & \langle u_a | v_l | \partial_{k_h} u_b\rangle \langle u_b | \Spin_i | u_a \rangle + \langle u_a | v_l | u_b \rangle \langle \partial_{k_h} u_b | \Spin_i | u_a \rangle \\
    &= \left[v_a^l \langle u_a | \partial_{k_h} u_b\rangle + \langle \partial_{k_l} u_a | \epsilon_a -H | \partial_{k_h} u_b\rangle \right] \Spin^i_{ba} \\
    & \hspace{1cm} + v_a^l \delta_{ab} \sum_{b'} \langle \partial_{k_h} u_a | u_{b'}\rangle \langle u_{b'} | \Spin_i | u_a \rangle. \nonumber 
\end{align}
Therefore, we obtain
\begin{align}
    \chi^{\mathrm{so(sta:intra)}}_{ij} &= -\frac{1}{V} \sum_{\bk} \sum_{a\equiv b} f'_a \left[ \frac{e}{2} (\bv_a \times \bA_{ab})_j +m_{ab}^j \right] \Spin^i_{ba} \nonumber \\
    & \quad +\frac{e}{2V} \sum_{\bk} \sum_{ab'} f'_a (\bv_a \times \bA_{ab'})_j \Spin^i_{b'a} \\
    &= \frac{e}{2V} \sum_{\bk} \sum_{a \not\equiv b} f'_a (\bv_a \times \bA_{ab})_j \Spin^i_{ba} \label{eq:intra-sta}\\
    & \quad -\frac{1}{V} \sum_{\bk} \sum_{a\equiv b} f'_a m_{ab}^j \Spin^i_{ba}. \nonumber
\end{align}

In the dynamical limit, the weight factor starts from $O(\bq^1)$,
\begin{align}
    \FF^{(1)h}_{ab}|_{\epsilon_a = \epsilon_b} &= \frac{v_a^h f'_a}{\Omega},
\end{align}
and hence the intraband contribution to the susceptibility becomes
\begin{align}
    \chi^{\mathrm{so(dyn:intra)}}_{ij} &= -\frac{ie}{2V} \epsilon_{jlh} \sum_{\bk} \sum_{a \equiv b} \frac{v_a^h f'_a}{\Omega} \MM_{ab}^{(0)il}. \label{eq:dyn-intra}
\end{align}
Here $\MM_{ab}^{(0)il}$ for $\epsilon_a = \epsilon_b$ reads
\begin{align}
    \langle u_a | v_l | u_b \rangle \langle u_b | \Spin_i | u_a \rangle &= v_a^l \delta_{ab} \Spin_{aa}^i.
\end{align}
Using this form, the right-hand side of Eq.~(\ref{eq:dyn-intra}) contains the factor $v_a^h v_a^l$,
which is symmetric in $l \leftrightarrow h$ and vanishes under the antisymmetrization.
Therefore, the intraband part for the dynamical susceptibility vanishes,
\begin{align}
    \chi^{\mathrm{so(dyn:intra)}}_{ij} &= 0. \label{eq:intra-dyn}
\end{align}

\subsubsection{Interband contribution}
The difference in the static and dynamical limits does not appear in the interband contribution.
The expansion of $\MM$ reads
\begin{align}
    \MM_{ab}^{(0)il} &= \langle u_a | v_l | u_b \rangle \langle u_b | \Spin_i | u_a \rangle \\
    &= - (\epsilon_a - \epsilon_b) \langle u_a | \partial_{k_l} u_b\rangle \Spin^i_{ba}, \\
    \MM_{ab}^{(1)ilh} &= \tfrac{1}{2} \langle u_a | \partial_{k_h} v_l | u_b\rangle \langle u_b | \Spin_i | u_a \rangle \\
    & \quad + \langle u_a | v_l | \partial_{k_h} u_b\rangle \langle u_b | \Spin_i | u_a \rangle \nonumber \\
    & \quad + \langle u_a | v_l | u_b\rangle \langle \partial_{k_h} u_b | \Spin_i | u_a \rangle \nonumber \\
    &= \tfrac{1}{2} \langle u_a | \partial_{k_l}\partial_{k_h} H | u_b\rangle \Spin^i_{ba} \label{eq:M1-expansion} \\
    & \quad + v_a^l \langle u_a | \partial_{k_h} u_b\rangle \Spin^i_{ba} \nonumber \\
    & \quad + \langle \partial_{k_l} u_a | \epsilon_a - H | \partial_{k_h} u_b\rangle \Spin^i_{ba} \nonumber \\
    & \quad + \langle \partial_{k_l} u_a | \epsilon_a-\epsilon_b | u_b\rangle \sum_c \langle \partial_{k_h} u_b | u_c \rangle \Spin^i_{ca} . \nonumber
\end{align}
By multiplying the weight factor, we are left with
\begin{align}
    \FF_{ab}^{(1)h} \MM_{ab}^{(0)il} &= v_b^h \left[f'_b -\frac{f_a - f_b}{\epsilon_a - \epsilon_b} \right] \langle u_a | \partial_{k_l} u_b\rangle \Spin^i_{ba}, \\
    \FF_{ab}^{(0)} \MM_{ab}^{(1)ilh} &\approx \frac{f_a - f_b}{\epsilon_a - \epsilon_b}\Spin^i_{ba} \label{eq:F0-M1} \\
    & \quad \times \left[ v_a^l \langle u_a | \partial_{k_h} u_b\rangle + \langle \partial_{k_l} u_a | \epsilon_a - H | \partial_{k_h} u_b\rangle \right] \nonumber \\
    & \quad + (f_a-f_b) \langle \partial_{k_l} u_a | u_b \rangle \sum_c \langle \partial_{k_h} u_b | u_c \rangle \Spin^i_{ca}. \nonumber
\end{align}
    Since the first term in Eq.~(\ref{eq:M1-expansion}) is symmetric in $l \leftrightarrow h$
    and does not contribute to the susceptibility,
    we have omitted its contribution to the right-hand side of Eq.~(\ref{eq:F0-M1}).
By identifying them with the geometric quantities term by term,
the interband contribution to the susceptibility reads
\begin{align}
    &\chi^{\mathrm{so(inter)}}_{ij} \nonumber \\
    &= -\frac{ie}{2V} \epsilon_{jlh} \sum_{\bk} \sum_{a\not\equiv b} \left[ \FF_{ab}^{(1)h} \MM_{ab}^{(0)il} + \FF_{ab}^{(0)} \MM_{ab}^{(1)ilh} \right]\\
    &= -\frac{e}{2V} \epsilon_{jlh} \sum_{\bk}\sum_{a\not \equiv b} \left[f'_b -\frac{f_a - f_b}{\epsilon_a - \epsilon_b} \right] v_b^h A_{ab}^l \Spin^i_{ba} \\
    & \quad -\frac{e}{2V} \epsilon_{jlh}\sum_{\bk}\sum_{a\not \equiv b} \frac{f_a - f_b}{\epsilon_a - \epsilon_b} v_a^l A_{ab}^h \Spin^i_{ba} \nonumber \\
    & \quad -\frac{1}{V} \sum_{\bk}\sum_{a\not \equiv b} \frac{f_a - f_b}{\epsilon_a - \epsilon_b} \left[ m_{ab}^j + \frac{e}{4}(\epsilon_a-\epsilon_b) \Omega_{ab}^j \right] \Spin^i_{ba} \nonumber \\
    & \quad -\frac{ie}{2V} \epsilon_{jlh} \sum_{\bk}\sum_{abc} (f_a-f_b) \langle \partial_{k_l} u_a | u_b \rangle \langle \partial_{k_h} u_b | u_c \rangle \Spin^i_{ca}, \nonumber
\end{align}
where the condition $a \not \equiv b$ in the last line is omitted
due to the factor $(f_a-f_b)$, which vanishes for $a \equiv b$.
By further processing these terms, we obtain
\begin{align}
    &\chi^{\mathrm{so(inter)}}_{ij}= \frac{e}{2V} \sum_{\bk} \sum_{a\not\equiv b} f'_a (\bv_a \times \bA_{ba})_j \Spin^i_{ab} \\
    & \quad -\frac{e}{2V}\sum_{\bk} \sum_{a\not\equiv b} \frac{f_a - f_b}{\epsilon_a - \epsilon_b} \left[ (\bv_a \times \bA_{ba})_j \Spin^i_{ab} + (\bv_a \times \bA_{ab})_j \Spin^i_{ba} \right] \nonumber \\
    & \quad -\frac{1}{V} \sum_{\bk}\sum_{a\not \equiv b} \frac{f_a - f_b}{\epsilon_a - \epsilon_b} \left[ m_{ab}^j + \frac{e}{4}(\epsilon_a-\epsilon_b) \Omega_{ab}^j \right] \Spin^i_{ba} \nonumber \\
    & \quad +\frac{ie}{2V} \epsilon_{jlh} \sum_{\bk}\sum_{abc} f_a \langle \partial_{k_l} u_a | u_b \rangle \langle u_b | \partial_{k_h} u_c \rangle \Spin^i_{ca} \nonumber \\
    & \quad -\frac{ie}{2V} \epsilon_{jlh} \sum_{\bk}\sum_{abc} f_b \langle \partial_{k_h} u_b | u_c \rangle \langle u_c | \Spin_i | u_a \rangle \langle u_a | \partial_{k_l} u_b \rangle \nonumber
\end{align}
\begin{align}
    \hspace{1cm} &= \frac{e}{2V} \sum_{\bk} \sum_{a\not\equiv b} f'_a (\bv_a \times \bA_{ba})_j \Spin^i_{ab} \\
    & \quad -\frac{e}{V}\sum_{\bk} \sum_{a\not\equiv b} \frac{f_a - f_b}{\epsilon_a - \epsilon_b} \mathrm{Re} \left[ (\bv_a \times \bA_{ab})_j \Spin^i_{ba} \right] \nonumber \\
    & \quad -\frac{1}{V} \sum_{\bk}\sum_{a\not \equiv b} \frac{f_a - f_b}{\epsilon_a - \epsilon_b} m_{ab}^j \Spin^i_{ba} \nonumber \\
    & \quad + \frac{e}{2V} \sum_{\bk} \sum_{ab} \frac{f_a+f_b}{2} \Omega_{ab}^j \Spin^i_{ba} + \frac{e}{2V} \sum_b f_b \Omega_{bb}^{(\Spin^i)j}. \nonumber
\end{align}
By adding the intraband contribution obtained in Eq.~(\ref{eq:intra-sta}) or (\ref{eq:intra-dyn}),
we obtain the full susceptibility classified by the geometric quantities,
as given in the main text.

%%% REFERENCES %%%1


\begin{thebibliography}{99}
\bibitem{Note-pseudospin}
Even if $\bsig$ in the Dirac Hamiltonian consists of pseudospin degrees of freedom,
our discussion is still valid,
by considering $\chi^{\mathrm{so}}$ as the response function
for the ``polarization'' (expectation value) of the pseudospin.
\end{thebibliography}
\end{document}